\documentclass[prd, aps, nofootinbib,twocolumn,floatfix,superscriptaddress]{revtex4-2}
\usepackage[T1]{fontenc}
\usepackage[utf8]{inputenc}
\usepackage[UKenglish]{babel}
\usepackage{amsmath,amssymb,mathtools, hyperref, tikz}

\usetikzlibrary{arrows.meta}

\newcommand{\pp}{\mathbf{p}}
\newcommand{\qq}{\mathbf{q}}
\newcommand{\zzero}{\mathbf{0}}
\newcommand{\ez}{\mathbf{e}_z}
\newcommand{\ex}{\mathbf{e}_x}
\newcommand{\ey}{\mathbf{e}_y}
\newcommand{\ei}{\mathbf{e}_i}

\newcommand{\appendixsection}{
	\setcounter{equation}{0}
	\setcounter{section}{0}
	\setcounter{figure}{0}
	\renewcommand{\theequation}{S\arabic{equation}}
	\renewcommand{\thefigure}{S\arabic{figure}}
	\onecolumngrid
	\vspace*{.7cm}
	\hrule
	\vspace*{.04cm}
	\hrule
	\begin{center}
		\vspace*{.4cm}
		{\bf \large Supplemental Material}
		\vspace*{.5cm}
	\end{center}
	\twocolumngrid
}

\begin{document}
\preprint{APS/123-QED}
\title{Toward $N$ to $N\pi$ matrix elements from lattice QCD}
\author{Lorenzo Barca}
\email{lorenzo.barca@desy.de}
\author{Gunnar Bali}
\email{gunnar.bali@ur.de}
\author{Sara Collins}
\email{sara.collins@ur.de}
\affiliation{
  Fakultät für Physik, Universität Regensburg,
  Universitätsstr.\ 31, 93053 Regensburg, Germany
}

\date{\today}

\begin{abstract}
  QCD matrix elements of axial and vector currents between nucleons
  are required for the Monte Carlo reconstruction of the energy of neutrinos
  that are detected in long baseline oscillation experiments in the
  quasielastic regime. The cleanest approach for determining the
  axial matrix elements is lattice QCD. However, the extraction
  of these from the corresponding correlation functions
  is complicated by very large excited state contributions, that are
  related to transitions from the nucleon to a nucleon-pion pair.
  In this pilot study with a pion mass $m_\pi = 429~\mathrm{MeV}$,
  we demonstrate for the first time that these contributions
  can be removed by including five-(anti)quark operators
  into the basis of interpolators used to create the nucleon.
  The same techniques will be needed to compute transition
  matrix elements between the nucleon and nucleon-pion
  scattering states that are relevant in the resonance production
  regime.\end{abstract}
\maketitle

\section{INTRODUCTION}
\vspace{-8pt}
The groundbreaking discovery of atmospheric and solar neutrino
oscillations more than two decades ago by the
Super-Kamiokande~\cite{Super-Kamiokande:1998kpq} and
SNO~\cite{SNO:2001kpb} experiments, respectively, required an
adjustment of the Standard Model to accommodate massive neutrinos.
The present generation of terrestrial long baseline neutrino
oscillation experiments, aimed at a more precise determination of the
neutrino masses and mixing parameters, NOvA~\cite{NOvA:2021nfi} and
T2K~\cite{T2K:2019bcf} as well as the future DUNE~\cite{DUNE:2020lwj}
experiment and the upgrade of T2K to the Hyper-Kamiokande
detector~\cite{Hyper-Kamiokande:2022smq} determine the fluxes of muon
and antimuon neutrinos via their interaction with nuclear targets in
a near and a far detector. The neutrino energies of the scattering
events are reconstructed via Monte Carlo event
generators~\cite{Andreopoulos:2009rq,GENIE:2021zuu}, which require
knowledge of the differential neutrino-nucleon cross section. For
neutrino energies below $1~\mathrm{GeV}$ this is dominated by
(quasi\nobreakdash-)elastic scattering, while from about
$400~\mathrm{MeV}$ onwards also resonance production with
nucleon-pion ($N\pi$) final states sets in \cite{Formaggio:2012cpf}.
Focusing on low energies, the cross section is proportional to the
square of a combination of nonperturbative nucleon vector and axial
matrix elements, which can be parameterized in terms of form factors.
The two vector form factors as functions of the squared four-momentum
transfer ($Q^2$) are sufficiently well known from experiment.
However, the two isovector axial form factors $G_A(Q^2)$ and
$\widetilde{G}_P(Q^2)$ are much less well constrained experimentally,
apart from $G_A$ in the forward limit ($Q^2=0$, i.e.~the axial charge
$g_A$~\cite{Markisch:2018ndu}) and
$\widetilde{G}_P(0.88\,m_{\mu}^2)=g_P^*$ at the muon capture point of
muonic hydrogen~\cite{MuCap:2012lei}.  Fortunately, these form factors
can be computed directly from QCD via lattice simulation. However,
there is a tension~\cite{Meyer:2022mix,Simons:2022ltq} between recent
lattice
results~\cite{RQCD:2019jai,Jang:2019vkm,Alexandrou:2020okk,Park:2021ypf,Djukanovic:2022wru}
and analyses of neutrino-deuteron scattering
experiments~\cite{Meyer:2016oeg}. Therefore, it is important to establish the
reliability of the lattice determinations. This requires the
investigation of all systematics and, in particular, the one
associated with extracting the nucleon matrix elements from
correlation functions at finite Euclidean times.  The latter receive
contributions also from single- and multiparticle states with the
same quantum numbers as the nucleon (normally referred to as excited
states).  At zero momentum, the lowest excitations with positive
parity include $N\pi$ P-wave and $N\pi\pi$ S-wave scattering states,
whereas, at nonvanishing momentum, parity is not a good quantum
number and also $N\pi$ in an S-wave can contribute.  Towards small
pion masses, the mass gap between the ground state and the first
excitation decreases and the spectrum becomes more dense.  Bearing in
mind that the signal-to-noise ratio of correlation functions decreases
exponentially with the Euclidean time separations, it can be very
challenging to reliably extract nucleon matrix elements. In order to
control the leading excited state contributions to these nucleon to
nucleon form factors, we will, for the first time, explicitly
calculate matrix elements that are also related to nucleon to $N\pi$
transition form factors, which are required for a firm understanding
of the resonance production regime.

Reliable continuum limit results for the axial form factors should
reproduce the experimentally known values of $g_A$ and $g_P^*$ and
also be consistent with the partially conserved axial current (PCAC)
relation (also referred to as the axial Ward identity (AWI)), which
relates the axial form factors to the pseudoscalar form factor. In
many previous simulations $g_A$ was reproduced, however, $g_P^*$
(defined at $Q^2>0$) was found to be smaller than the experimental
value and also the AWI between form factors was significantly
violated~\cite{Bali:2014nma,Capitani:2017qpc,Gupta:2017dwj,Tsukamoto:2017fnm,Alexandrou:2020okk}.
Since the AWI was found to be satisfied on the level of the
correlation functions in the continuum limit~\cite{Bali:2018qus}, the
inconsistency had to be related to the difficulty of isolating the
ground state contribution when extracting the 
form factors~\cite{Bali:2018qus,Jang:2019vkm,RQCD:2019jai}.  While the
interpolator that is used to create the nucleon was found to have
little overlap with excited states, as evidenced by analyses of
two-point functions, transition matrix elements between different
states, contributing to the spectral decomposition of the three-point
function, appeared to be enhanced.  Indeed, in chiral perturbation
theory (ChPT) the axial and pseudoscalar currents directly couple to
the pion. Regarding the pseudoscalar current or the time-component of
the axial current, $N$ to $N\pi$ transitions can contribute substantially 
to the three-point functions~\cite{Bar:2018xyi,Bar:2019gfx,Bar:2019igf} 
(see also Refs.~\cite{Tiburzi:2015tta,Hansen:2016qoz}). 
At a small but nonvanishing momentum, the leading such contribution increases 
in proportion to the ratio of the nucleon mass over the pion energy,
$m_N/E_{\pi}$~\cite{Bar:2018xyi,Bar:2019gfx}.  These terms
were taken into account in recent analyses of the Euclidean 
time dependence of lattice correlation functions, where form factors were
obtained, that are consistent with the
AWI~\cite{RQCD:2019jai,Jang:2019vkm,Park:2021ypf}.  However, the size
of the excited state contamination, found in these analyses, is quite
large in some channels for the Euclidean times that are accessible at
present. A more reliable approach would be to construct optimized
interpolators to minimize the dominant ($N\pi$) excited state
contributions.

In this work, we take into account directly the $N\pi$ contribution by
constructing nucleon-pion-like interpolators [$(qqq) (\bar{q}q)$ with
the quarks $q\in\{u,d\}$] $\mathrm{O}_{5q}$, and computing the
associated two-point and, for the first time, three-point correlation
functions between the standard three-quark nucleon interpolator
$\mathrm{O}_{3q}$ and $\mathrm{O}_{5q}$. Using this basis, that has
good overlap both with the nucleon ground state and the lowest lying
$N\pi$ state, nucleon to nucleon three-point functions can be
constructed with minimized $N\pi$ contributions, enhancing the
reliability of the extraction of the nucleon matrix elements.  As
mentioned above, this is the first step towards determining nucleon to
nucleon-pion matrix elements, associated with neutrino scattering in
the resonance production
regime~\cite{Simons:2022ltq,Ruso:2022qes,Formaggio:2012cpf}.  In this
pilot study we carry out the analysis for a single unphysical pion mass
$m_{\pi}=429~\mathrm{MeV}$.  It turns out that even at this relatively
large value the $N\pi$ contribution is very significant and that
this can effectively be removed with our approach. 
We expect this method to work even better at the physical pion mass:
ChPT becomes more reliable as the pion mass is reduced, 
and the tree-level $N\pi$ contribution is even more dominant, 
which is consistent with the observations made in Refs.~\cite{RQCD:2019jai, Jang:2019vkm, Bar:2018xyi, Bar:2019igf, Bar:2019gfx}.

\section{DEFINITION OF THE FORM FACTORS}
\vspace{-8pt}
We define local isovector pseudoscalar and axial currents,
$\mathcal{P}=\bar{d}\gamma_5u$ and
$\mathcal{A}_{\mu}=\bar{d}\gamma_{\mu}\gamma_5u$, respectively.  The
Lorentz decompositions into form factors of the respective matrix
elements read as
\begin{align}
	\label{lorentz_pseudoscalar_nme}
	\langle n_{\pp'} | \mathcal{P}| p_{\pp} \rangle &= 
	\bar{u}_{\pp'}  G_P(Q^2)\gamma_5 u_{\pp},
	\\
	\label{lorentz_axial_nme}
	\!\!\langle n_{\pp'} | \mathcal{A}_\mu| p_{\pp} \rangle &=
	\bar{u}_{\pp'}\!\!\left[\gamma_\mu G_A(Q^2) + \frac{q_\mu}{2m_N} \widetilde{G}_P(Q^2)\right]\!\!\gamma_5 u_{\pp},
\end{align}
where we assume isospin symmetry (i.e.\ $m_N=m_p=m_n$ and
$m_{\ell}=m_u=m_d$ for the quark masses), $u_{\pp}$ is the spinor of a
nucleon with three-momentum $\pp$, $q_{\mu}=p'_{\mu}-p_{\mu}$ is the
four-momentum transfer and $Q^2=-q_{\mu}q^{\mu}$.  Note that the above
decomposition of the axial matrix element does not hold if the two
states differ in their mass, e.g., if a nucleon is on the right-hand
side and a $N\pi$ on the left.  The AWI $\partial_\mu \mathcal{A}^\mu
= 2im_\ell \mathcal{P}$ implies the relation between form factors,
\(m_N G_A(Q^2) = m_\ell G_P(Q^2) + (Q^2/4m_N) \widetilde{G}_P(Q^2)\),
which is exact in the continuum limit but will be affected by moderate
discretization effects at our lattice spacing $a\approx
0.098$~fm~\cite{Bali:2018qus,RQCD:2019jai}.  In addition, the pion
pole dominance (PPD) assumption gives the approximate relation,
\(\widetilde{G}_P(Q^2) \approx 4 m_N^2 G_A(Q^2)/(m_\pi^2+Q^2).\) While
this only holds exactly for $m_{\pi}=0$, in Ref.~\cite{RQCD:2019jai} it was
found to hold within uncertainties of 1\%--2\% at the physical point
in the continuum limit, with violations of less than 3\% up to
$m_{\pi}\approx 420$~MeV. Deviations from these relations can be
quantified in terms of the differences from unity of the combinations
\begin{align}
  r_{\rm{PCAC}} &= \frac{4m_Nm_\ell G_P (Q^2)+ Q^2 \widetilde{G}_P(Q^2)}{4m_N^2G_A(Q^2)},  \label{eq_ratioPCAC}
\\
  r_{\rm{PPD}} &= \frac{(m_\pi^2+Q^2) \widetilde{G}_P(Q^2)}{4 m_N^2 G_A(Q^2)}.
  \label{eq_ratioPPD}
\end{align}
\section{ANALYSIS} \vspace{-8pt}
We construct the matrices of two- and three-point correlation
functions~(see the Supplemental Material),
\begin{align}
	\label{gevp_matrix_2pt}
	C_{2pt}(\pp, t)_{ij} &= \left\langle\mathrm{O}_i(\pp, t) ~\bar{\mathrm{O}}_j(\pp, 0)\right\rangle,
	\\
	\label{gevp_matrix_3pt}
	C^{\mathcal{J}}_{3pt}(\pp', t; \qq, \tau)_{ij} &= \left\langle\mathrm{O}_i(\pp', t) ~\mathcal{J}(\qq, \tau) ~\bar{\mathrm{O}}_j(\pp, 0)\right\rangle,
\end{align}
where we indicate the three-momentum transfer in the argument of the
local current $\mathcal{J}\in\{\mathcal{P},\mathcal{A}_{\mu}\}$.  The
interpolators $\mathrm{O}_i\in\{\mathrm{O}_{3q}, \mathrm{O}_{5q}\}$
are projected onto the $G_1$ representation of the double cover of the
cubic group $^2O_h$ (or, for nonvanishing momentum, the relevant
little group)~\cite{Gockeler:2012yj,Lang:2016hnn,Prelovsek:2016iyo},
corresponding to spin and helicity $1/2$ in the continuum, as well as
to definite momentum and isospin. For instance, $I_3=-1/2$ corresponds
to $\mathrm{O}_{3q}\sim n$ and $\mathrm{O}_{5q} \sim
\sqrt{1/3} \, n\pi^0 - \sqrt{2/3} \, p\pi^-$.  The
Wick contractions of the correlation functions are evaluated using the
sequential method~\cite{Maiani:1987by} for quark-line connected
topologies, while the stochastic
``one-end-trick''~\cite{Sommer:1994gg,Foster:1998wu,Foster:1999oet} is
used for disconnected diagrams.

For the results shown here, about 200 propagators~(see the Supplemental Material) for each
of the six source-sink separations have been computed on 800 gauge
configurations. In view of the computational cost, we carry out the analysis
on a single coordinated lattice simulations (CLS~\cite{Bruno:2014jqa})
ensemble (A653, see Ref.~\cite{RQCD:2022xux}) with the spatial volume
$L^3=(24a)^3$, employing $N_f=3$ nonperturbatively improved Wilson
fermions with a lattice spacing $a\approx 0.098~\mathrm{fm}$ and the
pion mass $m_\pi= 429~\mathrm{MeV}$.  The best results were obtained,
using extended (smeared) quark fields in the nucleon and pion
interpolators. (For details on the smearing, see appendix C.1 and
Table~15 of Ref.~\cite{RQCD:2022xux}).

\begin{figure}[htb]
  \includegraphics[width=\linewidth]{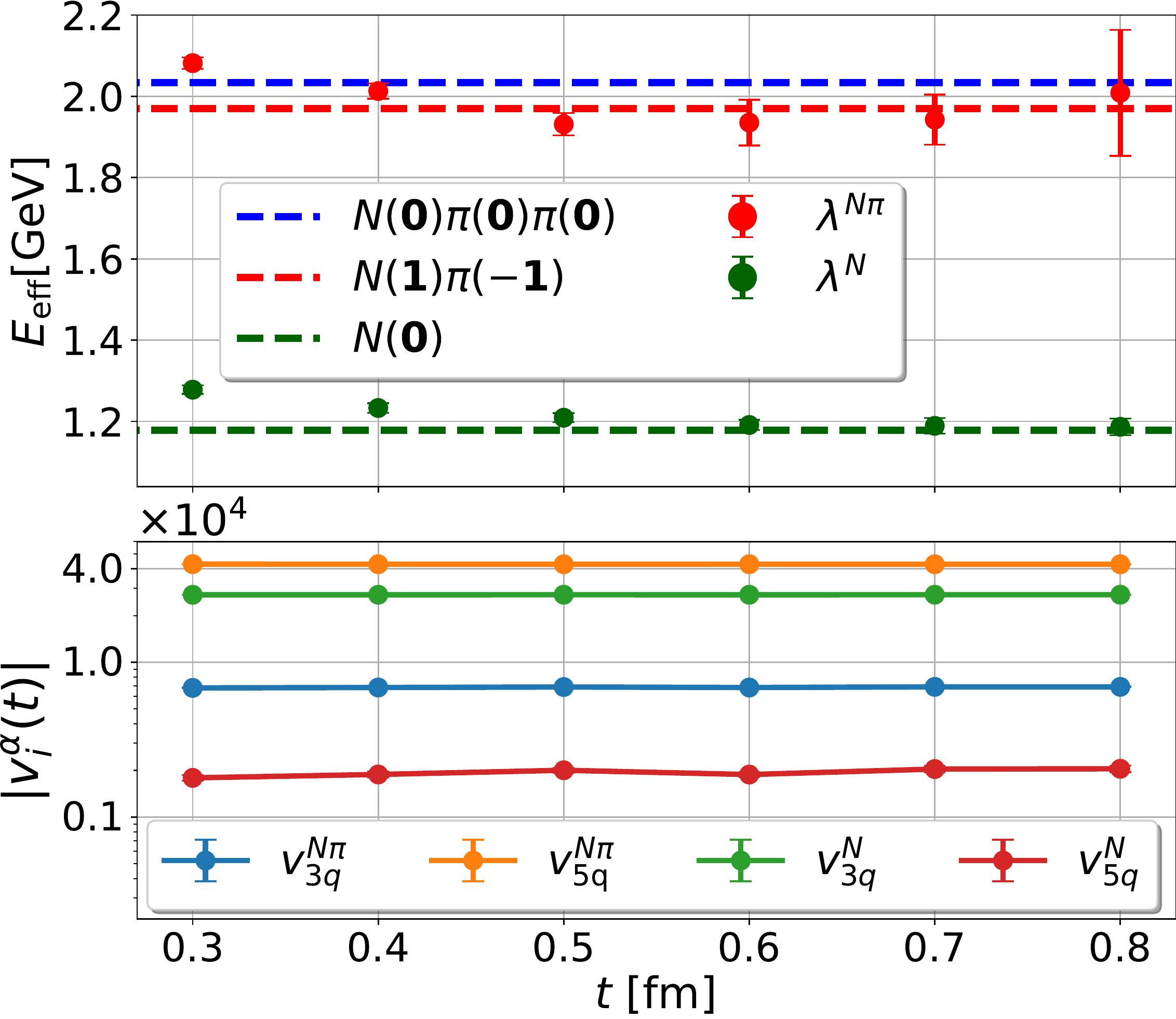}
  \caption{Top: the effective energies obtained solving the GEVP for $\pp=\zzero$ 
    and $t_0=0.2~\mathrm{fm}$, compared with the nucleon mass and the
    lowest energy levels of noninteracting $N\pi$ P- and $N\pi\pi$ S-waves.
    Bottom: the moduli of the corresponding generalized
    eigenvector components.\label{fig:gevp_lambda_mom0}}
\end{figure}

We extract the generalized eigenvalue and eigenvector matrices
$\Lambda(\pp, t;t_0)=\mathrm{diag}\left(\lambda^1(\pp, t; t_0),
\lambda^2(\pp, t; t_0) \right)$ and $V(\pp, t, t_0)=\left(v^1(\pp, t;
t_0), v^2(\pp, t; t_0)\right)$, respectively, by solving the
generalized eigenvalue problem
(GEVP)~\cite{Berg:1982hf,Michael:1985ne,Luscher:1990ck,Blossier:2009kd}
for the matrix of two-point functions, \(C_{2pt}(\pp, t) V(\pp, t,
t_0) = C_{2pt}(\pp, t_0) V(\pp, t, t_0) \Lambda(\pp, t, t_0)\), for
fixed reference times $t_0$, where we employ the normalization
$v^{\alpha\intercal}C_{2pt}(t_0)v^{\alpha}=1$. For large times $t$ the
eigenvalues will decay exponentially with the energy of the state:
$\lambda^{\alpha}(\pp, t;t_0) \rightarrow d^{\alpha}(\pp; t_0)
e^{-E^\alpha(\pp) (t-t_0)}$, where $d^{\alpha}\lesssim 1$.

The effective energies $E_{\rm
  eff}^{\alpha}(t)=a^{-1}\ln[\lambda^{\alpha}(t)/\lambda^{\alpha}(t+a)]$
are shown in Fig.~\ref{fig:gevp_lambda_mom0} for $\pp=\zzero$ and
$t_0=0.2~\mathrm{fm}$. The lowest energy coincides with the nucleon
mass on this ensemble~\cite{RQCD:2022xux}, while the second level is
close to the sum of the nucleon and pion energies for the lowest
P-wave momentum combination. Therefore, we will identify $N$ with
$\alpha=1$ and $N\pi$ with $\alpha=2$.  Note that the eigenvectors are
very stable in $t$ and that the contribution of $\mathrm{O}_{5q}$
(subscript $i=2$) to the nucleon state is suppressed by more than $1$
order of magnitude relative to $\mathrm{O}_{3q}$. Nevertheless, as we
will see, the impact on three-point functions can be significant.  We
also solve the GEVP for moving frames, in particular for $\pp =
\ez\eqqcolon\frac{2\pi}{L}(0, 0, 1)$ ($|\ez|\approx 530~\mathrm{MeV}$).
Regarding $\mathrm{O}_{5q}$, we
consider the combinations
$\mathrm{O}_{3q}(\ez)\mathrm{O}_{\bar{q}q}(\zzero)$ and
$\mathrm{O}_{3q}(\zzero)\mathrm{O}_{\bar{q}q}(\ez)$, with
$\mathrm{O}_{\bar{q}q}$ being a pion interpolator.  Solving the GEVP,
in both cases we find the effective energy of the second eigenvalue
for $t>0.5~\mathrm{fm}$ to be consistent with the $N(\ez)\pi(\zzero)$
and $N(\zzero)\pi(\ez)$ noninteracting energies, respectively.

Considering these results, we employ the generalized eigenvectors for
$t_0=0.2~\mathrm{fm}$ and $t=0.5~\mathrm{fm}$ to construct the
GEVP-optimized correlation functions,
\begin{align}	
	\label{projected_2pt}
	\mathcal{C}_{2pt}(\pp, t)^{\alpha} 
	&=  v_i^\alpha (\pp) C_{2pt}(\pp, t)_{ij} v_j^\alpha (\pp),
	\\
	\label{projected_3pt}
	\mathcal{C}_{3pt}^{\mathcal{J}}(\pp', t; \qq, \tau)^{\alpha\beta} 
	&= v_i^\alpha (\pp')C^{\mathcal{J}}_{3pt}(\pp', t; \qq, \tau)_{ij} v_j^\beta (\pp ),
\end{align}
where $\alpha,\beta\in\{N,N\pi\}$. Note that here we only present
results for $\alpha=\beta=N$ and we neglect the $i=j=2$ element ($5q$
to $5q$) of $C_{3pt}^{\mathcal{J}}$ that, in this case, is suppressed
by the second power of the small eigenvector component $v_{5q}^N$. In
addition, from ChPT we would only expect nondiagonal elements of the
matrices of correlators to be enhanced.  The nucleon matrix elements
of interest are then extracted by forming the GEVP ratios~\cite{Bulava:2011yz,Owen:2012ts,Dragos:2016rtx}
\begin{align}\nonumber
  R_{\mathcal{J}}(\pp', t; \qq, \tau)
  &=
  \frac{\mathcal{C}_{3pt}^{\mathcal{J}}(\pp', t; \qq, \tau)^{NN} }{\mathcal{C}_{2pt}(\pp', t)^{N} }\\\nonumber\times&
  \sqrt{
    \frac{
      \mathcal{C}_{2pt}(\pp', \tau)^{N}
      \mathcal{C}_{2pt}(\pp', t)^{N}
      \mathcal{C}_{2pt}(\pp, t-\tau)^{N}
    }
	 {
	   \mathcal{C}_{2pt}(\pp, \tau)^{N}
	   \mathcal{C}_{2pt}(\pp, t)^{N}
	   \mathcal{C}_{2pt}(\pp', t-\tau)^{N}
	 }
  }
  \\
  &\propto
  \langle N_{\pp'} | \mathcal{J}| N_{\pp} \rangle
  \qquad(t\gg \tau \gg 0),
  \label{gevp_improved_ratio}
\end{align}
where excited state contributions of the type $N \rightarrow N\pi$ and
$N\pi\rightarrow N$ are explicitly removed. Forming the same ratio for
the usual two- and three-point functions, $C_{2pt,11}$ and
$C_{3pt,11}^{\mathcal{J}}$, will give the same result in the limit of
large $t$ and $\tau$. Any time dependence observed for these ratios
(GEVP-improved or not) is an indication of remaining excited state
contamination.

\section{RESULTS IN THE FORWARD LIMIT} \vspace{-8pt}
For $\pp'=\pp$ the combination under the square root in
Eq.~\eqref{gevp_improved_ratio} cancels. We consider two kinematic
combinations: $\pp'=\pp=\zzero$ and $\pp'=\pp=\ez$. Regarding the rest
frame, the three-point functions vanish due to parity for
$\mathcal{J}=\mathcal{A}_4$ and $\mathcal{J}=\mathcal{P}$, while
$\mathcal{J}=\mathcal{A}_i$ (with the spin projected in the
$i$ direction) at large Euclidean time separations gives the axial
charge $g_A$.  Contamination from the coupling to $N\pi$ states
exists, however, only as a loop effect in ChPT. Indeed, even when
using the standard ratio, for $t>0.8~\mathrm{fm}$ the data near
$\tau=t/2$ show no time dependence within their errors. Fitting the
(unimproved) ratio for $1.15~\mathrm{fm}<t<1.4~\mathrm{fm}$, we find
$g_A=1.156(7)$ at our unphysical pion mass.

\begin{figure}[htb]
  \includegraphics[width=\linewidth]{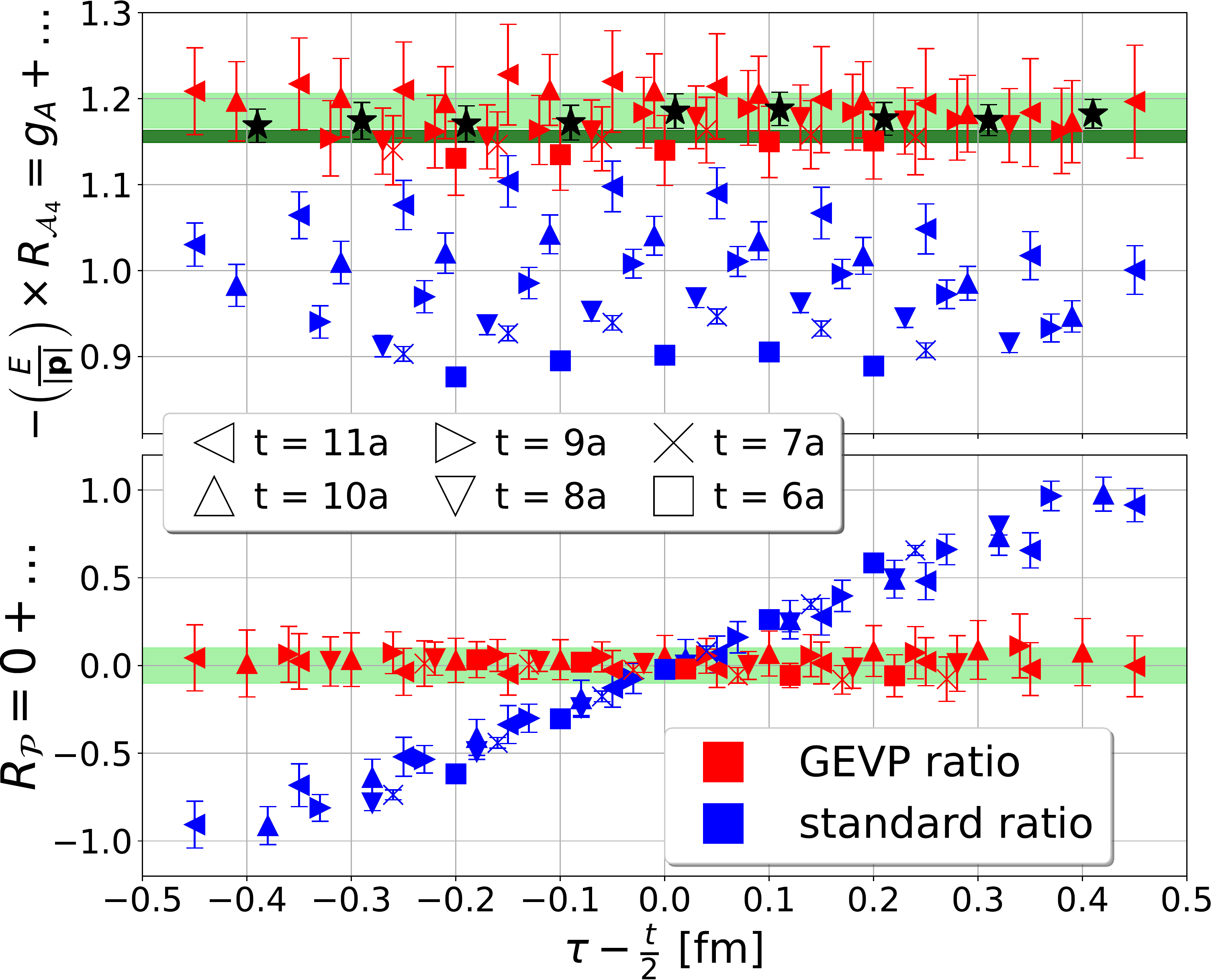}
  \caption{Comparison of the ratios in Eq.~\eqref{gevp_improved_ratio} in
    the forward limit, constructed from the standard and
    GEVP-optimized correlation functions, with $\pp=\pp'=\ez$
    as a function of the current insertion time $\tau$ for a
    number of different source-sink separations $t$, where $a\approx
    0.098~\mathrm{fm}$.  Top: renormalized ratio for the time
    component of the axial current.  Also shown are the standard
    ratios for $\mathcal{A}_z$ at $\pp=\ez$ and $t=10a$ (black stars).
    The dark and light green bands correspond to $g_A$, extracted from
    this ratio at $\pp=\zzero$ and at $\pp=\ez$, respectively, using
    different $t$.  Bottom: the same for the
    pseudoscalar current. The green band highlights the expected
    result.
    \label{fig:comparison_plots_standard_gevp_ratios_forward_limit}}
\end{figure}

Also in the moving frame the $N\pi$ contributions to $\mathcal{A}_i$
only appear as loop effects, and we find that the corresponding
standard ratio is almost constant (black stars in
Fig.~\ref{fig:comparison_plots_standard_gevp_ratios_forward_limit}).
However, regarding $\mathcal{A}_4$ and $\mathcal{P}$, $N$ to $N\pi$
transitions appear at tree-level and are enhanced by one power of
$m_N/E_{\pi}$, relative to the $N$ to $N$ matrix elements of
interest. The ratio $R_{\mathcal{A}_4}$ will be proportional to $g_A$
at large times too; however, using the $\mathrm{O}_{3q}$
interpolators, we find substantial excited state contamination, which
is indicated by its strong dependence on the source-sink separation,
see the blue symbols in the upper panel of
Fig.~\ref{fig:comparison_plots_standard_gevp_ratios_forward_limit}.  A
difference between the ratios for $\mathcal{A}_i$ and $\mathcal{A}_4$
at $\pp'=\pp\neq\zzero$ was also observed, e.g.,
in Ref.~\cite{Liang:2016fgy}, using standard interpolators.  In contrast,
the GEVP-improved ratios (red symbols) already agree for
$t>0.6~\mathrm{fm}$ with the value extracted from the standard ratios
for $\mathcal{A}_z$ obtained at $\pp\neq\zzero$ (light green band) and
at $\pp=\zzero$ (dark green band).
A similar, dramatic reduction of the excited state contamination is
observed for the pseudoscalar current for the GEVP-optimized
correlation functions. For this current, due to charge conjugation
symmetry, diagonal matrix elements vanish. Therefore, the deviation
from zero of the standard ratio in the lower panel of
Fig.~\ref{fig:comparison_plots_standard_gevp_ratios_forward_limit} is
entirely an excited state effect, which is removed within the present
errors for the GEVP-optimized ratio. This demonstrates that the
dominant contribution is from $N$ to $N\pi$ transitions, which is
consistent with the tree-level ChPT expectation.

\section{RESULTS FOR NONVANISHING MOMENTUM TRANSFER} \vspace{-8pt}
When determining the axial and pseudoscalar form factors, the excited state
contamination is prominent for correlation functions involving
the currents $\mathcal{J}\in \{
\mathcal{A}_4,\mathcal{P}\}$~\cite{Bar:2018xyi,Bar:2019gfx,RQCD:2019jai,Bar:2019igf},
that can transfer the momentum to a pion at tree-level in ChPT.  
\begin{figure}[htb]
	\includegraphics[height=6.9cm, width=\linewidth]{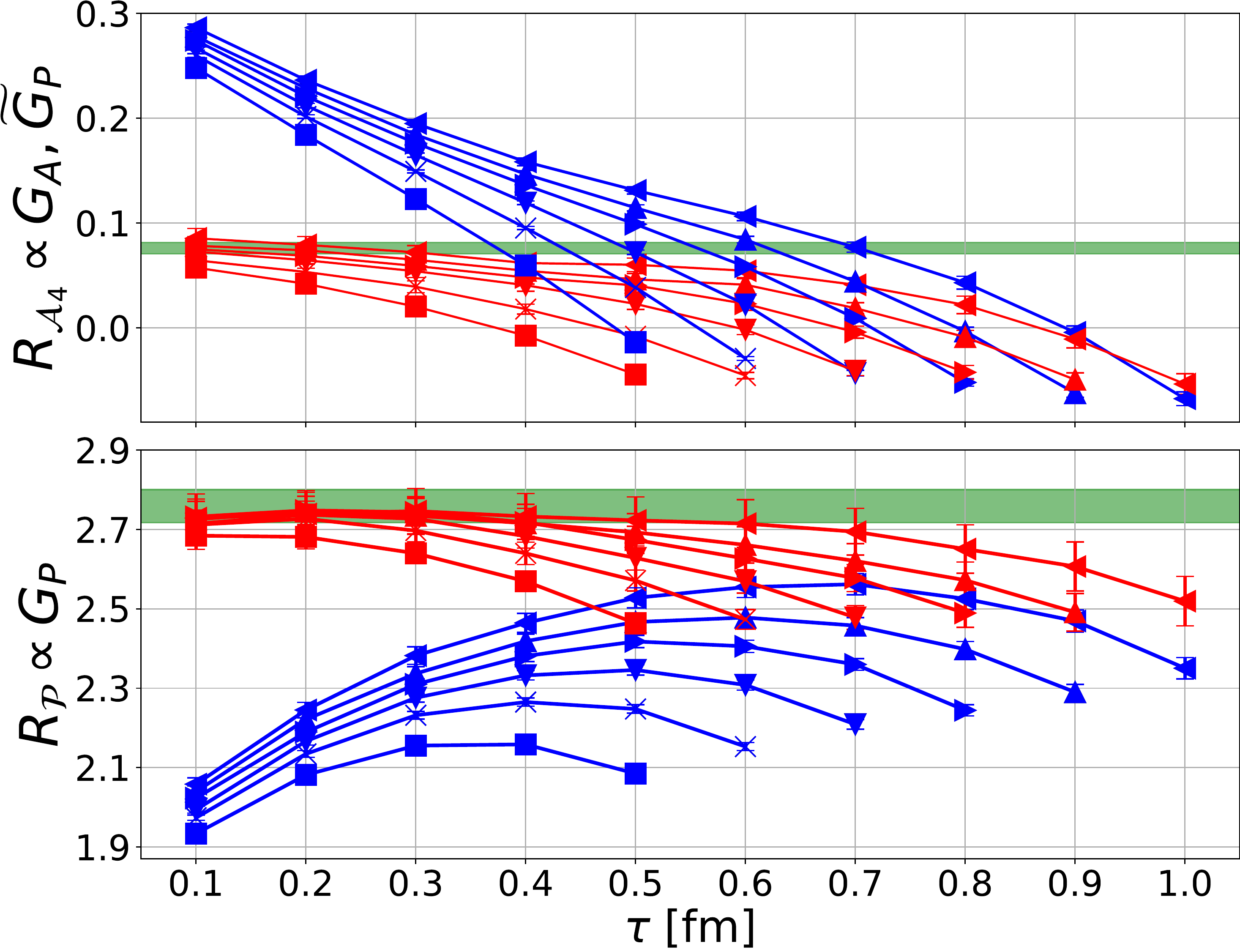}
	\caption{The same as in
		Fig.~\ref{fig:comparison_plots_standard_gevp_ratios_forward_limit}
		(but as a function of the insertion time $\tau$) for
		$\pp'=\zzero\neq\pp=-\ez$, corresponding to $Q^2 \approx
		0.3~\mathrm{GeV}^2$.  The green bands represent results of a
		simultaneous fit to the shown GEVP-optimized ratios and the
		standard ratio for $\mathcal{A}_z$.
		\label{fig:comparison_plots_standard_gevp_ratios}}
\end{figure}
This contribution, which is proportional to
$m_N/E_{\pi}$~\cite{Bar:2018xyi,Bar:2019gfx},
is largest at small momentum transfer.  Therefore, we
consider these two currents and set $\qq=\ez$ to a single unit of
lattice momentum ($|\qq|\approx 530~\mathrm{MeV}$).  With
$\pp'=\zzero$ and $\pp=-\qq$, this corresponds to $Q^2\approx
0.3~\mathrm{GeV}^2$.  Our results for the two ratios for the standard
and optimized correlation functions are shown in
Fig.~\ref{fig:comparison_plots_standard_gevp_ratios}. 
Clearly, the time dependence
is much reduced for the GEVP-optimized results: at
the source, excited states are effectively removed; however, at the
sink (that is at rest) there are clearly residual effects from higher
excitations.
The ratios at large times (green bands) are proportional to the
respective matrix elements which, using the
decompositions~\eqref{lorentz_pseudoscalar_nme}
and~\eqref{lorentz_axial_nme}, are related to a linear combination of
$G_A$ and $\widetilde{G}_P$ for $\mathcal{A}_4$ and 
$G_P$ for $\mathcal{P}$, respectively.

The axial form factor $G_A= 0.91 \pm 0.01$ is extracted from the standard 
correlation functions with $\mathcal{A}_i$ and $\ei\perp \qq$.
These show ground state dominance within our range of $t$ and $\tau$.
Indeed, the large tree-level $N\pi$ ChPT diagrams do not contribute to 
this channel and only $N\pi$ loop diagrams appear \cite{Bar:2018xyi}. 
The ground state matrix element is proportional only to $G_A$ and we use 
the value we extract as a prior in fits to the other channels. 
We fit the GEVP-optimized ratios for the pseudoscalar and
the temporal axial currents simultaneously to constants plus 
exponentials $\propto e^{-\Delta E(t-\tau)}$. 

\begin{figure}[htb]
	\includegraphics[width=\linewidth]{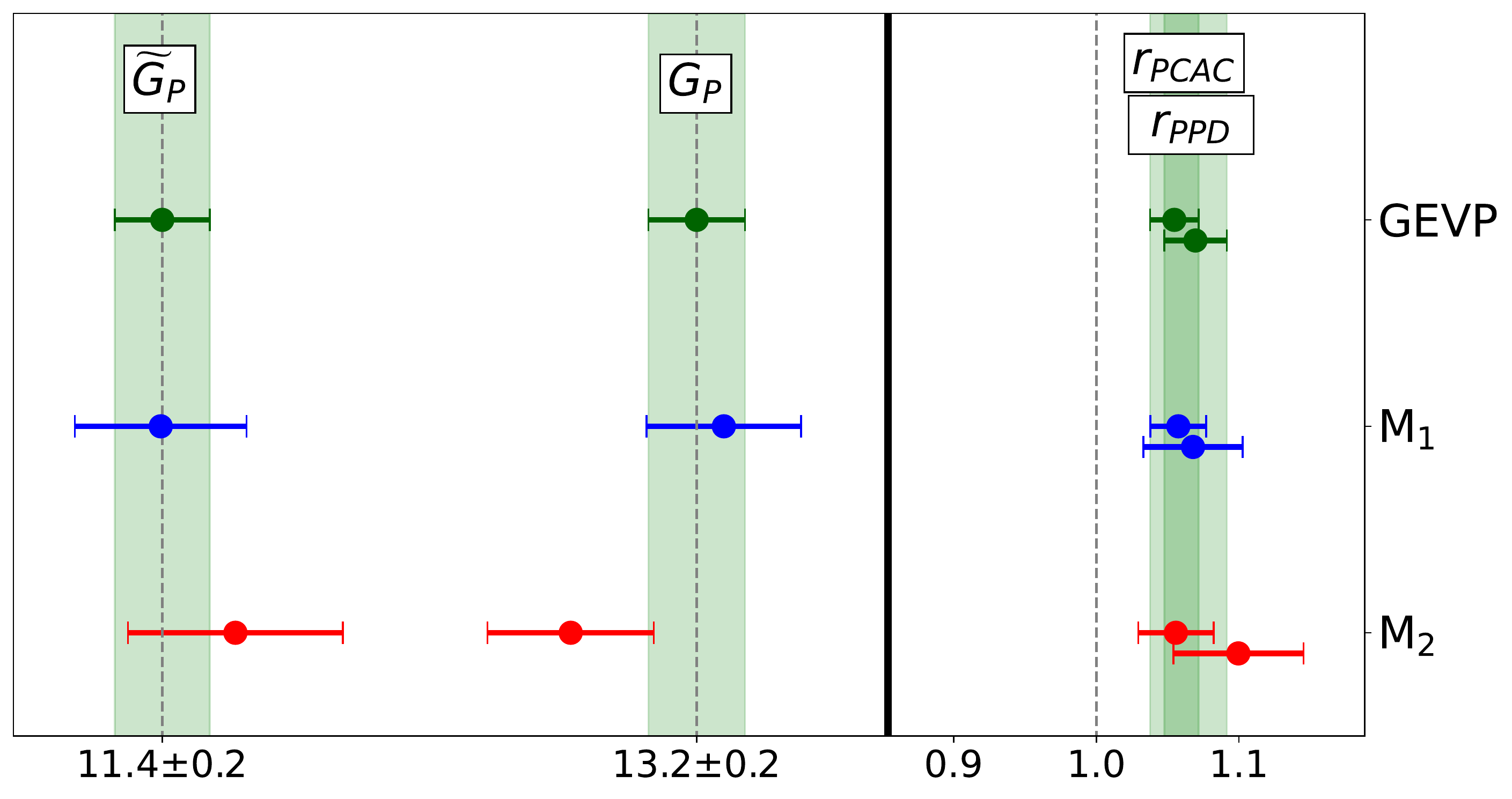}
	\caption{Results for the form factors and the PCAC and PPD
		ratios~\eqref{eq_ratioPCAC} and~\eqref{eq_ratioPPD} at $Q^2\approx
		0.3~\mathrm{GeV}^2$ from the GEVP-optimized correlators, in
		comparison to results obtained from the standard correlation
		functions, using the ChPT guided multistate analysis techniques
		of Ref.~\cite{RQCD:2019jai} ($\mathrm{M}_1$) and inspired by \cite{Jang:2019vkm}
		($\mathrm{M}_2$).
		\label{fig:comparison_different_methods}}
\end{figure}

We find $\Delta E\approx 2m_{\pi}$ for the gap
between the nucleon ground state and this first excitation. The
resulting matrix elements then give the pseudoscalar and induced
pseudoscalar form factors at $Q^2\approx 0.3~\mathrm{GeV}^2$, the
latter after subtracting the $G_A$ contribution. The results for 
$G_P$ and $\widetilde{G}_P$ as well as for the PCAC and PPD
ratios of Eqs.~\eqref{eq_ratioPCAC} and~\eqref{eq_ratioPPD} are shown
in Fig.~\ref{fig:comparison_different_methods}. We also include
results that are obtained using the ChPT guided methods
of Ref.~\cite{RQCD:2019jai} ($\mathrm{M}_1$) and a simultaneous fit 
to the channels $\mathcal{A}_4$, $\mathcal{P}$ and $\mathcal{A}_i$ with 
$\qq = \ei$, inspired by \cite{Jang:2019vkm} ($\mathrm{M}_2$).
In spite of the large excited state contributions, the
results using modern multistate analysis techniques agree within
errors with the GEVP results, at least at $m_{\pi}=429~\mathrm{MeV}$.
At this single lattice spacing, the ratio $r_{\mathrm{PCAC}}$ somewhat
differs from $1$.

\section{CONCLUSIONS}\vspace{-8pt}
Given the current tension~\cite{Meyer:2022mix,Simons:2022ltq} between
results for the axial form factor obtained from lattice QCD and from
reanalyses of historical neutrino-deuteron scattering
experiments~\cite{Meyer:2016oeg}, it is important to rigorously
investigate the systematics associated with the lattice approach.  The
PCAC relation between form factors has only recently been verified in
some studies~\cite{RQCD:2019jai,Jang:2019vkm,Park:2021ypf} and this
provides an important cross check.  We have shown that the very large
excited state contributions encountered can be removed by including
$N\pi$-type interpolators. This confirms ChPT expectations, even at
our relatively large pion mass, and supports assumptions made in
recent determinations of the axial 
form factor~\cite{RQCD:2019jai,Jang:2019vkm,Alexandrou:2020okk,Park:2021ypf,Djukanovic:2022wru}.
In the near future, we will repeat the study at a smaller pion mass
where excited state contributions are even larger, with the aim of
also determining $N$ to $N\pi$ matrix elements that are relevant for
the scattering of neutrinos with energies larger than
$400~\mathrm{MeV}$.
\newpage
\begin{acknowledgments}\vspace{-8pt}
The authors thank M.~Göckeler, M.~Padmanath, S.~Prelovsek, P.~Wein and
T.~Wurm for discussion.  L.B. is grateful to G.~Herdoiza for engaging
discussions during his secondment to IFT Madrid and to A.~Kronfeld,
M.~Wagman, W.~Detmold and P.~Shanahan for fruitful discussions and
their hospitality at Fermilab and MIT.  This project has received
funding from the European Union's Horizon 2020 research and innovation
program under the Marie Skłodowska Grant Agreement
No.~813942 (ITN EuroPLEx) and the EU Grant Agreement No.~824093
(STRONG~2020). Support from the MISTI Global Seed Fund MIT-Germany -
University of Regensburg project ``Quark and gluon structure of the
proton'' is gratefully acknowledged.  The authors thank the
\href{https://www.gauss-centre.eu}{Gauss Centre for Supercomputing
  (GCS)} for providing computing time through the
\href{http://www.john-von-neumann-institut.de}{John von Neumann
  Institute for Computing (NIC)} on the Booster partition of the
supercomputer JURECA~\cite{jureca} at
\href{http://www.fz-juelich.de/ias/jsc/}{J\"ulich Supercomputing
  Centre (JSC)}. G. C. S. is the alliance of the three national
supercomputing centres HLRS (Universität Stuttgart), JSC
(Forschungszentrum Jülich), and LRZ (Bayerische Akademie der
Wissenschaften), funded by the BMBF and the German State Ministries
for Research of Baden\nobreakdash-Württemberg (MWK), Bayern (StMWFK)
and Nordrhein\nobreakdash-Westfalen (MIWF). Simulations were also
performed on the QPACE~3 computer of SFB/TR\nobreakdash-55, using an
adapted version of the {\sc Chroma}~\cite{Edwards:2004sx} software
package.
\end{acknowledgments}

\newpage
\appendixsection
\section*{Construction of the two- and three-point functions}\vspace{-0pt}
In order to employ the generalized eigenvalue (and eigenstate)
approach, the following two-point correlation functions (see
Eq.~(5))
\begin{align}
	\label{N2N_2pt}
	&\langle
	\mathrm{O}_{3q}(\mathbf{p}, t)
	~\bar{\mathrm{O}}_{3q}(\mathbf{p}, 0)
	\rangle,
	\\
	\label{N2Npi_2pt}
	&\langle
	\mathrm{O}_{5q}(\mathbf{p}, t)
	~\bar{\mathrm{O}}_{3q}(\mathbf{p}, 0)
	\rangle,
	\\
	\label{Npi2Npi_2pt}
	&\langle
	\mathrm{O}_{5q}(\mathbf{p}, t)
	~\bar{\mathrm{O}}_{5q}(\mathbf{p}, 0)
	\rangle
\end{align}
and three-point correlation functions (see Eq.~(6))
\begin{align}
	\label{std_3pt}
	&\langle
	\mathrm{O}_{3q}(\mathbf{p}', t)
	~\mathcal{J}(\mathbf{q}, \tau)
	~\bar{\mathrm{O}}_{3q}(\mathbf{p}, 0)
	\rangle,      \\
	\label{N2Npi_3pt}
	&\langle
	\mathrm{O}_{5q}(\mathbf{p}', t)
	~\mathcal{J}(\mathbf{q}, \tau)
	~\bar{\mathrm{O}}_{3q}(\mathbf{p}, 0)
	\rangle
\end{align}
need to be evaluated.  In these expressions, $\mathrm{O}_{3q}$
represents a nucleon-like interpolating operator with a $3$-quark
$qqq$ structure and $\mathrm{O}_{5q}$ is nucleon-pion-like with a
$(qqq)(\bar{q}q)$-structure.

We first discuss the construction of the three-point functions, where
we consider transitions from an $I=I_3=1/2$ state (e.g., the proton
$p$) via a charged current $\mathcal{J}=\bar{d}\Gamma u$ to an
$I=-I_3=1/2$ state (e.g., the neutron, $n$),
i.e.\ $\bar{\mathrm{O}}_{3q}$ has the flavour structure
$\bar{u}\bar{u}\bar{d}\sim \bar{p}$, whereas $\mathrm{O}_{3q}$
corresponds to $udd\sim n$. To project $\mathrm{O}_{5q}$ onto
$I=-I_3=1/2$, the combination $\sqrt{1/3} \, n\pi^0 -
\sqrt{2/3} \, p\pi^-$ must be formed, where
$\pi^-\sim\bar{u}d$ and $\pi^0\sim 1/\sqrt{2}(\bar{u}u-\bar{d}{d})$.
Like $\mathrm{O}_{3q}$ also $\mathrm{O}_{5q}$ must be projected onto
the lattice irreducible representation $G_1$, see, e.g.,
Refs.~\cite{Prelovsek:2016iyo, Lang:2016hnn}.  In the rest frame, for
the spin-up component, we form the combination
\begin{align}
	\mathrm{O}_{5q}^{G_1, \uparrow}(\zzero)
	=& \nonumber
	~
	\mathrm{O}_{3q}(-\ex)
	\mathrm{O}_{\bar{q}q}(\ex)
	-
	\mathrm{O}_{3q}(\ex)
	\mathrm{O}_{\bar{q}q}(-\ex)
	\\
	&
	\nonumber
	-
	i\mathrm{O}_{3q}(-\ey)
	\mathrm{O}_{\bar{q}q}(\ey)
	+i
	\mathrm{O}_{3q}(\ey)
	\mathrm{O}_{\bar{q}q}(-\ey)
	\\
	&
	+
	~
	\mathrm{O}_{3q}(-\ez)
	\mathrm{O}_{\bar{q}q}(\ez)
	-
	\mathrm{O}_{3q}(\ez)
	\mathrm{O}_{\bar{q}q}(-\ez),
\end{align}
where $\ei$ corresponds to one unit of lattice momentum in the
$i$-direction and $\mathrm{O}_{\bar{q}q}$ is a pion interpolator.
Regarding a moving frame with $\pp'=\ez$, we employ two
combinations that, in the continuum limit, will project on the helicity $+1/2$:
\begin{align}
	\mathrm{O}_{5q}^{G_1, \uparrow}(\ez)^1
	&=
	~
	\mathrm{O}^{\uparrow}_{3q}(\ez)
	\mathrm{O}_{\bar{q}q}(\zzero),\label{eq:nomom}
	\\
	\mathrm{O}_{5q}^{G_1, \uparrow}(\ez)^2
	&=
	~
	\mathrm{O}^{\uparrow}_{3q}(\zzero)
	\mathrm{O}_{\bar{q}q}(\ez).\label{eq:moment}
\end{align}
Following tree-level ChPT, in the forward limit ($\qq=\zzero$), in the
moving frame ($\pp=\pp'=\ez$), Eq.~\eqref{eq:nomom} is the relevant
interpolator, while for off-forward kinematics ($\pp'=\ez$,
$\pp=\zzero$) Eq.~\eqref{eq:moment} is used.

\begin{figure}[hbt]
	\includegraphics[width=\columnwidth]{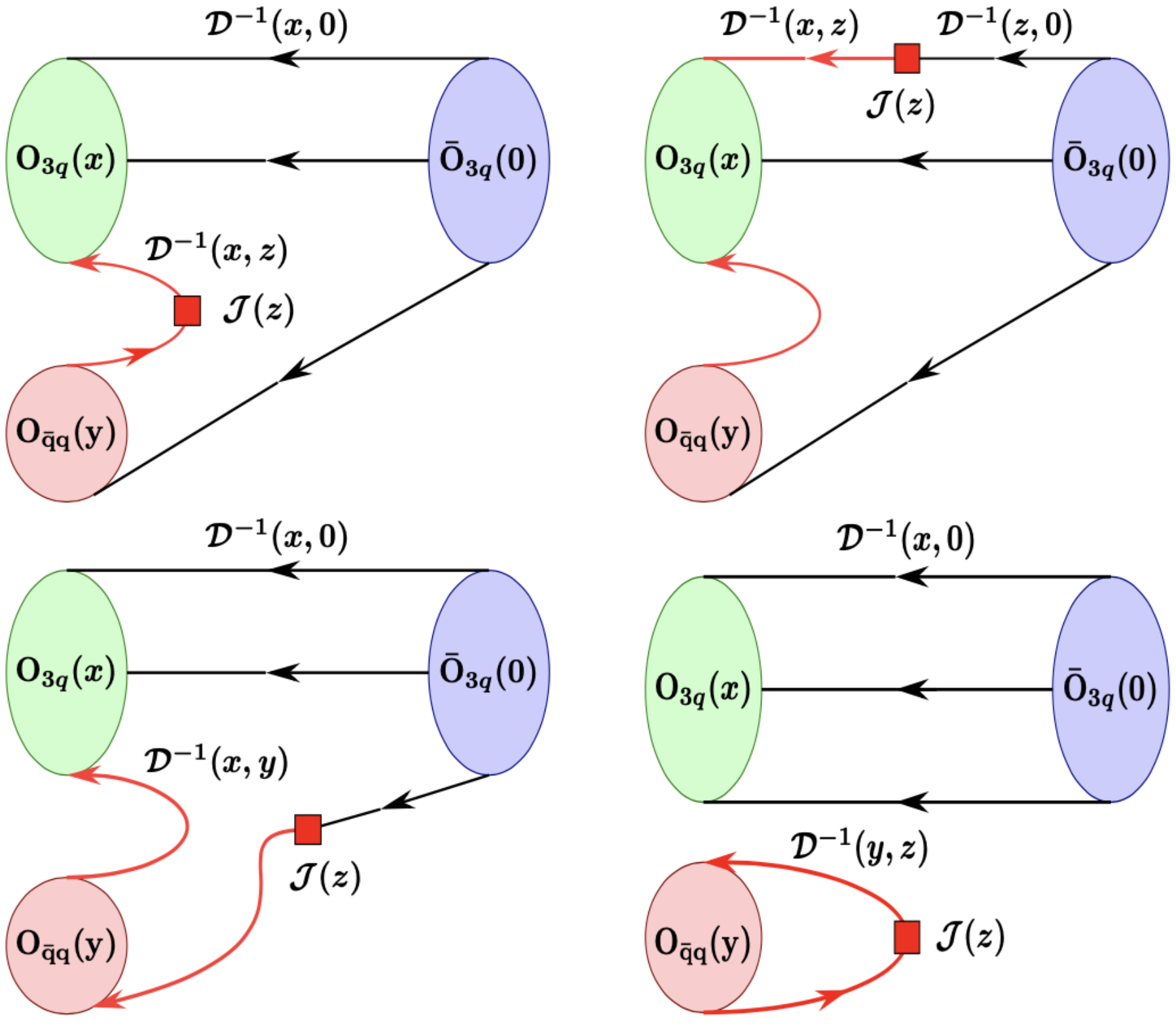}
	\caption{Topologies in position space that appear after performing
		the Wick contractions of the three-point correlation functions
		$\langle \mathrm{O}_{5q}(x, y) ~\mathcal{J}(z)
		~\bar{\mathrm{O}}_{3q}(0) \rangle$.  The three black lines
		represent point-to-all propagators, the two red lines correspond
		to all-to-all propagators and the red square indicates the current
		insertion at $z=(\mathbf{z},\tau)$.  The $\mathrm{O}_{5q}(x, y)$
		interpolator contains a $qqq$-structure ($\mathrm{O}_{3q}(x)$) at
		the spacetime position $x=(\mathbf{x}, t)$ and a
		$\bar{q}q$-structure ($\mathrm{O}_{\bar{q}q}(y)$) at
		$y=(\mathbf{y}, t)$.
		\label{fig:diagrams_nJnpi}
	}
\end{figure}

Performing the Wick contractions, we encounter the four different
quark-line diagram topologies shown in Fig.~\ref{fig:diagrams_nJnpi}.
Each topology represents a number of different Wick contractions.
Note that for the charged current we are interested in, the

disconnected diagram does not contribute to the transition
$p\rightarrow n\pi^0$, whereas all four diagrams contribute to
$p\rightarrow p\pi^-$.  
In each diagram there are two all-to-all
propagators (red lines).  

The quark-line disconnected diagram on the
bottom-right is computed, combining a point-to-all nucleon two-point
function with a stochastically estimated meson two-point function,
using the ``one-end-trick''~\cite{Sommer:1994gg,Foster:1998wu,Foster:1999oet}.
For the quark-line connected diagrams, we use the sequential source
method~\cite{Maiani:1987by} to compute the all-to-all propagators.
The Wick contractions for the two-point functions
Eq.~\eqref{Npi2Npi_2pt} have the same topologies and are computed in a
similar way, replacing the current by a smeared pion interpolating
operator at $\tau=0$.

Most computer time is spent on the quark-line connected diagrams,
where we employ the sequential-source method twice for each
combination of momentum projections at $y$ and at $x$\footnote{ We use
	$\mathcal{D}^{-1}(x,z)\gamma_5\mathcal{D}^{-1}(z,x)\gamma_5$ (and the
	same for $\mathcal{D}^{-1}(x,y)$). Some building blocks are
	shared between different diagrams.} as well as for each spin
polarization of $\mathrm{O}_{3q}$.  The disconnected diagram is much
less expensive.  The total cost for each nucleon source position and
fixed source-sink separation on a given configuration is equivalent to
the computation of about $200$ propagators. This can be
compared to three propagators for the standard three-point function,
when setting $\pp'=\zzero$ and considering the unpolarized case and 
one polarization. Including $\pp'=\pm\ei$ (as we do here) increases 
this cost five-fold (15 propagators) and there would be an additional 
factor if further polarizations were evaluated. Therefore, the addition 
of the $\mathrm{O}_{5q}$ interpolators increases the computational 
complexity by about one order of magnitude.

\bibliography{bibliography}

\begin{thebibliography}{49}%
\makeatletter
\providecommand \@ifxundefined [1]{%
 \@ifx{#1\undefined}
}%
\providecommand \@ifnum [1]{%
 \ifnum #1\expandafter \@firstoftwo
 \else \expandafter \@secondoftwo
 \fi
}%
\providecommand \@ifx [1]{%
 \ifx #1\expandafter \@firstoftwo
 \else \expandafter \@secondoftwo
 \fi
}%
\providecommand \natexlab [1]{#1}%
\providecommand \enquote  [1]{``#1''}%
\providecommand \bibnamefont  [1]{#1}%
\providecommand \bibfnamefont [1]{#1}%
\providecommand \citenamefont [1]{#1}%
\providecommand \href@noop [0]{\@secondoftwo}%
\providecommand \href [0]{\begingroup \@sanitize@url \@href}%
\providecommand \@href[1]{\@@startlink{#1}\@@href}%
\providecommand \@@href[1]{\endgroup#1\@@endlink}%
\providecommand \@sanitize@url [0]{\catcode `\\12\catcode `\$12\catcode
  `\&12\catcode `\#12\catcode `\^12\catcode `\_12\catcode `\%12\relax}%
\providecommand \@@startlink[1]{}%
\providecommand \@@endlink[0]{}%
\providecommand \url  [0]{\begingroup\@sanitize@url \@url }%
\providecommand \@url [1]{\endgroup\@href {#1}{\urlprefix }}%
\providecommand \urlprefix  [0]{URL }%
\providecommand \Eprint [0]{\href }%
\providecommand \doibase [0]{https://doi.org/}%
\providecommand \selectlanguage [0]{\@gobble}%
\providecommand \bibinfo  [0]{\@secondoftwo}%
\providecommand \bibfield  [0]{\@secondoftwo}%
\providecommand \translation [1]{[#1]}%
\providecommand \BibitemOpen [0]{}%
\providecommand \bibitemStop [0]{}%
\providecommand \bibitemNoStop [0]{.\EOS\space}%
\providecommand \EOS [0]{\spacefactor3000\relax}%
\providecommand \BibitemShut  [1]{\csname bibitem#1\endcsname}%
\let\auto@bib@innerbib\@empty
\bibitem [{\citenamefont {Fukuda}\ \emph {et~al.}(1998)\citenamefont {Fukuda}
  \emph {et~al.}}]{Super-Kamiokande:1998kpq}%
  \BibitemOpen
  \bibfield  {author} {\bibinfo {author} {\bibfnamefont {Y.}~\bibnamefont
  {Fukuda}} \emph {et~al.} (\bibinfo {collaboration} {Super-Kamiokande}),\
  }\bibfield  {title} {\bibinfo {title} {{Evidence for oscillation of
  atmospheric neutrinos}},\ }\href
  {https://doi.org/10.1103/PhysRevLett.81.1562} {\bibfield  {journal} {\bibinfo
   {journal} {Phys. Rev. Lett.}\ }\textbf {\bibinfo {volume} {81}},\ \bibinfo
  {pages} {1562} (\bibinfo {year} {1998})},\ \Eprint
  {https://arxiv.org/abs/hep-ex/9807003} {arXiv:hep-ex/9807003} \BibitemShut
  {NoStop}%
\bibitem [{\citenamefont {Ahmad}\ \emph {et~al.}(2001)\citenamefont {Ahmad}
  \emph {et~al.}}]{SNO:2001kpb}%
  \BibitemOpen
  \bibfield  {author} {\bibinfo {author} {\bibfnamefont {Q.~R.}\ \bibnamefont
  {Ahmad}} \emph {et~al.} (\bibinfo {collaboration} {SNO}),\ }\bibfield
  {title} {\bibinfo {title} {{Measurement of the rate of $\nu_e+d \to p+p+e^-$
  interactions produced by $^8$B solar neutrinos at the Sudbury Neutrino
  Observatory}},\ }\href {https://doi.org/10.1103/PhysRevLett.87.071301}
  {\bibfield  {journal} {\bibinfo  {journal} {Phys. Rev. Lett.}\ }\textbf
  {\bibinfo {volume} {87}},\ \bibinfo {pages} {071301} (\bibinfo {year}
  {2001})},\ \Eprint {https://arxiv.org/abs/nucl-ex/0106015}
  {arXiv:nucl-ex/0106015} \BibitemShut {NoStop}%
\bibitem [{\citenamefont {Acero}\ \emph {et~al.}(2022)\citenamefont {Acero}
  \emph {et~al.}}]{NOvA:2021nfi}%
  \BibitemOpen
  \bibfield  {author} {\bibinfo {author} {\bibfnamefont {M.~A.}\ \bibnamefont
  {Acero}} \emph {et~al.} (\bibinfo {collaboration} {NOvA}),\ }\bibfield
  {title} {\bibinfo {title} {{Improved measurement of neutrino oscillation
  parameters by the NOvA experiment}},\ }\href
  {https://doi.org/10.1103/PhysRevD.106.032004} {\bibfield  {journal} {\bibinfo
   {journal} {Phys. Rev. D}\ }\textbf {\bibinfo {volume} {106}},\ \bibinfo
  {pages} {032004} (\bibinfo {year} {2022})},\ \Eprint
  {https://arxiv.org/abs/2108.08219} {arXiv:2108.08219 [hep-ex]} \BibitemShut
  {NoStop}%
\bibitem [{\citenamefont {Abe}\ \emph {et~al.}(2020)\citenamefont {Abe} \emph
  {et~al.}}]{T2K:2019bcf}%
  \BibitemOpen
  \bibfield  {author} {\bibinfo {author} {\bibfnamefont {K.}~\bibnamefont
  {Abe}} \emph {et~al.} (\bibinfo {collaboration} {T2K}),\ }\bibfield  {title}
  {\bibinfo {title} {{Constraint on the matter\textendash{}antimatter
  symmetry-violating phase in neutrino oscillations}},\ }\href
  {https://doi.org/10.1038/s41586-020-2177-0} {\bibfield  {journal} {\bibinfo
  {journal} {Nature}\ }\textbf {\bibinfo {volume} {580}},\ \bibinfo {pages}
  {339} (\bibinfo {year} {2020})},\ \bibinfo {note} {[Erratum: Nature 583, E16
  (2020)]},\ \Eprint {https://arxiv.org/abs/1910.03887} {arXiv:1910.03887
  [hep-ex]} \BibitemShut {NoStop}%
\bibitem [{\citenamefont {Abi}\ \emph {et~al.}(2020)\citenamefont {Abi} \emph
  {et~al.}}]{DUNE:2020lwj}%
  \BibitemOpen
  \bibfield  {author} {\bibinfo {author} {\bibfnamefont {B.}~\bibnamefont
  {Abi}} \emph {et~al.} (\bibinfo {collaboration} {DUNE}),\ }\bibfield  {title}
  {\bibinfo {title} {{Deep Underground Neutrino Experiment (DUNE), far detector
  technical design report, volume I introduction to DUNE}},\ }\href
  {https://doi.org/10.1088/1748-0221/15/08/T08008} {\bibfield  {journal}
  {\bibinfo  {journal} {JINST}\ }\textbf {\bibinfo {volume} {15}}\bibfield
  {number} {\bibinfo  {number} { (08)},\ \bibinfo {pages} {T08008}},\ }\Eprint
  {https://arxiv.org/abs/2002.02967} {arXiv:2002.02967 [physics.ins-det]}
  \BibitemShut {NoStop}%
\bibitem [{\citenamefont {Bian}\ \emph {et~al.}(2022)\citenamefont {Bian} \emph
  {et~al.}}]{Hyper-Kamiokande:2022smq}%
  \BibitemOpen
  \bibfield  {author} {\bibinfo {author} {\bibfnamefont {J.}~\bibnamefont
  {Bian}} \emph {et~al.} (\bibinfo {collaboration} {Hyper-Kamiokande}),\
  }\bibfield  {title} {\bibinfo {title} {{Hyper-Kamiokande experiment: a
  Snowmass white paper}},\ }in\ \href@noop {} {\emph {\bibinfo {booktitle}
  {{2022 Snowmass Summer Study}}}}\ (\bibinfo {year} {2022})\ \Eprint
  {https://arxiv.org/abs/2203.02029} {arXiv:2203.02029 [hep-ex]} \BibitemShut
  {NoStop}%
\bibitem [{\citenamefont {Andreopoulos}\ \emph {et~al.}(2010)\citenamefont
  {Andreopoulos} \emph {et~al.}}]{Andreopoulos:2009rq}%
  \BibitemOpen
  \bibfield  {author} {\bibinfo {author} {\bibfnamefont {C.}~\bibnamefont
  {Andreopoulos}} \emph {et~al.} (\bibinfo {collaboration} {GENIE}),\
  }\bibfield  {title} {\bibinfo {title} {{The GENIE neutrino Monte Carlo
  generator}},\ }\href {https://doi.org/10.1016/j.nima.2009.12.009} {\bibfield
  {journal} {\bibinfo  {journal} {Nucl. Instrum. Meth. A}\ }\textbf {\bibinfo
  {volume} {614}},\ \bibinfo {pages} {87} (\bibinfo {year} {2010})},\ \Eprint
  {https://arxiv.org/abs/0905.2517} {arXiv:0905.2517 [hep-ph]} \BibitemShut
  {NoStop}%
\bibitem [{\citenamefont {Tena-Vidal}\ \emph {et~al.}(2021)\citenamefont
  {Tena-Vidal} \emph {et~al.}}]{GENIE:2021zuu}%
  \BibitemOpen
  \bibfield  {author} {\bibinfo {author} {\bibfnamefont {J.}~\bibnamefont
  {Tena-Vidal}} \emph {et~al.} (\bibinfo {collaboration} {GENIE}),\ }\bibfield
  {title} {\bibinfo {title} {{Neutrino-nucleon cross-section model tuning in
  GENIE v3}},\ }\href {https://doi.org/10.1103/PhysRevD.104.072009} {\bibfield
  {journal} {\bibinfo  {journal} {Phys. Rev. D}\ }\textbf {\bibinfo {volume}
  {104}},\ \bibinfo {pages} {072009} (\bibinfo {year} {2021})},\ \Eprint
  {https://arxiv.org/abs/2104.09179} {arXiv:2104.09179 [hep-ph]} \BibitemShut
  {NoStop}%
\bibitem [{\citenamefont {Formaggio}\ and\ \citenamefont
  {Zeller}(2012)}]{Formaggio:2012cpf}%
  \BibitemOpen
  \bibfield  {author} {\bibinfo {author} {\bibfnamefont {J.~A.}\ \bibnamefont
  {Formaggio}}\ and\ \bibinfo {author} {\bibfnamefont {G.~P.}\ \bibnamefont
  {Zeller}},\ }\bibfield  {title} {\bibinfo {title} {{From eV to EeV: neutrino
  cross sections across energy scales}},\ }\href
  {https://doi.org/10.1103/RevModPhys.84.1307} {\bibfield  {journal} {\bibinfo
  {journal} {Rev. Mod. Phys.}\ }\textbf {\bibinfo {volume} {84}},\ \bibinfo
  {pages} {1307} (\bibinfo {year} {2012})},\ \Eprint
  {https://arxiv.org/abs/1305.7513} {arXiv:1305.7513 [hep-ex]} \BibitemShut
  {NoStop}%
\bibitem [{\citenamefont {M\"arkisch}\ \emph {et~al.}(2019)\citenamefont
  {M\"arkisch} \emph {et~al.}}]{Markisch:2018ndu}%
  \BibitemOpen
  \bibfield  {author} {\bibinfo {author} {\bibfnamefont {B.}~\bibnamefont
  {M\"arkisch}} \emph {et~al.},\ }\bibfield  {title} {\bibinfo {title}
  {{Measurement of the weak axial-vector coupling constant in the decay of free
  neutrons using a pulsed cold neutron beam}},\ }\href
  {https://doi.org/10.1103/PhysRevLett.122.242501} {\bibfield  {journal}
  {\bibinfo  {journal} {Phys. Rev. Lett.}\ }\textbf {\bibinfo {volume} {122}},\
  \bibinfo {pages} {242501} (\bibinfo {year} {2019})},\ \Eprint
  {https://arxiv.org/abs/1812.04666} {arXiv:1812.04666 [nucl-ex]} \BibitemShut
  {NoStop}%
\bibitem [{\citenamefont {Andreev}\ \emph {et~al.}(2013)\citenamefont {Andreev}
  \emph {et~al.}}]{MuCap:2012lei}%
  \BibitemOpen
  \bibfield  {author} {\bibinfo {author} {\bibfnamefont {V.~A.}\ \bibnamefont
  {Andreev}} \emph {et~al.} (\bibinfo {collaboration} {MuCap}),\ }\bibfield
  {title} {\bibinfo {title} {{Measurement of muon capture on the proton to 1\%
  precision and determination of the pseudoscalar coupling $g_P$}},\ }\href
  {https://doi.org/10.1103/PhysRevLett.110.012504} {\bibfield  {journal}
  {\bibinfo  {journal} {Phys. Rev. Lett.}\ }\textbf {\bibinfo {volume} {110}},\
  \bibinfo {pages} {012504} (\bibinfo {year} {2013})},\ \Eprint
  {https://arxiv.org/abs/1210.6545} {arXiv:1210.6545 [nucl-ex]} \BibitemShut
  {NoStop}%
\bibitem [{\citenamefont {Meyer}\ \emph {et~al.}(2022)\citenamefont {Meyer},
  \citenamefont {Walker-Loud},\ and\ \citenamefont
  {Wilkinson}}]{Meyer:2022mix}%
  \BibitemOpen
  \bibfield  {author} {\bibinfo {author} {\bibfnamefont {A.~S.}\ \bibnamefont
  {Meyer}}, \bibinfo {author} {\bibfnamefont {A.}~\bibnamefont {Walker-Loud}},\
  and\ \bibinfo {author} {\bibfnamefont {C.}~\bibnamefont {Wilkinson}},\
  }\bibfield  {title} {\bibinfo {title} {{Status of lattice QCD determination
  of nucleon form factors and their relevance for the few-GeV neutrino
  program}},\ }\href {https://doi.org/10.1146/annurev-nucl-010622-120608}
  {\bibfield  {journal} {\bibinfo  {journal} {Annu. Rev. Nucl. Part. Sci.}\
  }\textbf {\bibinfo {volume} {72}},\ \bibinfo {pages} {205} (\bibinfo {year}
  {2022})},\ \Eprint {https://arxiv.org/abs/2201.01839} {arXiv:2201.01839
  [hep-lat]} \BibitemShut {NoStop}%
\bibitem [{\citenamefont {Simons}\ \emph {et~al.}(2022)\citenamefont {Simons},
  \citenamefont {Steinberg}, \citenamefont {Lovato}, \citenamefont {Meurice},
  \citenamefont {Rocco},\ and\ \citenamefont {Wagman}}]{Simons:2022ltq}%
  \BibitemOpen
  \bibfield  {author} {\bibinfo {author} {\bibfnamefont {D.}~\bibnamefont
  {Simons}}, \bibinfo {author} {\bibfnamefont {N.}~\bibnamefont {Steinberg}},
  \bibinfo {author} {\bibfnamefont {A.}~\bibnamefont {Lovato}}, \bibinfo
  {author} {\bibfnamefont {Y.}~\bibnamefont {Meurice}}, \bibinfo {author}
  {\bibfnamefont {N.}~\bibnamefont {Rocco}},\ and\ \bibinfo {author}
  {\bibfnamefont {M.}~\bibnamefont {Wagman}},\ }\bibfield  {title} {\bibinfo
  {title} {{Form factor and model dependence in neutrino-nucleus cross section
  predictions}},\ }\href@noop {} {\  (\bibinfo {year} {2022})},\ \Eprint
  {https://arxiv.org/abs/2210.02455} {arXiv:2210.02455 [hep-ph]} \BibitemShut
  {NoStop}%
\bibitem [{\citenamefont {Bali}\ \emph {et~al.}(2020)\citenamefont {Bali},
  \citenamefont {Barca}, \citenamefont {Collins}, \citenamefont {Gruber},
  \citenamefont {L\"offler}, \citenamefont {Sch\"afer}, \citenamefont
  {S\"oldner}, \citenamefont {Wein}, \citenamefont {Weish\"aupl},\ and\
  \citenamefont {Wurm}}]{RQCD:2019jai}%
  \BibitemOpen
  \bibfield  {author} {\bibinfo {author} {\bibfnamefont {G.~S.}\ \bibnamefont
  {Bali}}, \bibinfo {author} {\bibfnamefont {L.}~\bibnamefont {Barca}},
  \bibinfo {author} {\bibfnamefont {S.}~\bibnamefont {Collins}}, \bibinfo
  {author} {\bibfnamefont {M.}~\bibnamefont {Gruber}}, \bibinfo {author}
  {\bibfnamefont {M.}~\bibnamefont {L\"offler}}, \bibinfo {author}
  {\bibfnamefont {A.}~\bibnamefont {Sch\"afer}}, \bibinfo {author}
  {\bibfnamefont {W.}~\bibnamefont {S\"oldner}}, \bibinfo {author}
  {\bibfnamefont {P.}~\bibnamefont {Wein}}, \bibinfo {author} {\bibfnamefont
  {S.}~\bibnamefont {Weish\"aupl}},\ and\ \bibinfo {author} {\bibfnamefont
  {T.}~\bibnamefont {Wurm}} (\bibinfo {collaboration} {RQCD}),\ }\bibfield
  {title} {\bibinfo {title} {{Nucleon axial structure from lattice QCD}},\
  }\href {https://doi.org/10.1007/JHEP05(2020)126} {\bibfield  {journal}
  {\bibinfo  {journal} {J. High Energy Phys.}\ }\textbf {\bibinfo {volume}
  {05}},\ \bibinfo {pages} {126}},\ \Eprint {https://arxiv.org/abs/1911.13150}
  {arXiv:1911.13150 [hep-lat]} \BibitemShut {NoStop}%
\bibitem [{\citenamefont {Jang}\ \emph {et~al.}(2020)\citenamefont {Jang},
  \citenamefont {Gupta}, \citenamefont {Yoon},\ and\ \citenamefont
  {Bhattacharya}}]{Jang:2019vkm}%
  \BibitemOpen
  \bibfield  {author} {\bibinfo {author} {\bibfnamefont {Y.-C.}\ \bibnamefont
  {Jang}}, \bibinfo {author} {\bibfnamefont {R.}~\bibnamefont {Gupta}},
  \bibinfo {author} {\bibfnamefont {B.}~\bibnamefont {Yoon}},\ and\ \bibinfo
  {author} {\bibfnamefont {T.}~\bibnamefont {Bhattacharya}},\ }\bibfield
  {title} {\bibinfo {title} {{axial vector form factors from lattice QCD that
  satisfy the PCAC relation}},\ }\href
  {https://doi.org/10.1103/PhysRevLett.124.072002} {\bibfield  {journal}
  {\bibinfo  {journal} {Phys. Rev. Lett.}\ }\textbf {\bibinfo {volume} {124}},\
  \bibinfo {pages} {072002} (\bibinfo {year} {2020})},\ \Eprint
  {https://arxiv.org/abs/1905.06470} {arXiv:1905.06470 [hep-lat]} \BibitemShut
  {NoStop}%
\bibitem [{\citenamefont {Alexandrou}\ \emph {et~al.}(2021)\citenamefont
  {Alexandrou} \emph {et~al.}}]{Alexandrou:2020okk}%
  \BibitemOpen
  \bibfield  {author} {\bibinfo {author} {\bibfnamefont {C.}~\bibnamefont
  {Alexandrou}} \emph {et~al.} (\bibinfo {collaboration} {ETM}),\ }\bibfield
  {title} {\bibinfo {title} {{Nucleon axial and pseudoscalar form factors from
  lattice QCD at the physical point}},\ }\href
  {https://doi.org/10.1103/PhysRevD.103.034509} {\bibfield  {journal} {\bibinfo
   {journal} {Phys. Rev. D}\ }\textbf {\bibinfo {volume} {103}},\ \bibinfo
  {pages} {034509} (\bibinfo {year} {2021})},\ \Eprint
  {https://arxiv.org/abs/2011.13342} {arXiv:2011.13342 [hep-lat]} \BibitemShut
  {NoStop}%
\bibitem [{\citenamefont {Park}\ \emph {et~al.}(2022)\citenamefont {Park},
  \citenamefont {Gupta}, \citenamefont {Yoon}, \citenamefont {Mondal},
  \citenamefont {Bhattacharya}, \citenamefont {Jang}, \citenamefont {Jo\'o},\
  and\ \citenamefont {Winter}}]{Park:2021ypf}%
  \BibitemOpen
  \bibfield  {author} {\bibinfo {author} {\bibfnamefont {S.}~\bibnamefont
  {Park}}, \bibinfo {author} {\bibfnamefont {R.}~\bibnamefont {Gupta}},
  \bibinfo {author} {\bibfnamefont {B.}~\bibnamefont {Yoon}}, \bibinfo {author}
  {\bibfnamefont {S.}~\bibnamefont {Mondal}}, \bibinfo {author} {\bibfnamefont
  {T.}~\bibnamefont {Bhattacharya}}, \bibinfo {author} {\bibfnamefont {Y.-C.}\
  \bibnamefont {Jang}}, \bibinfo {author} {\bibfnamefont {B.}~\bibnamefont
  {Jo\'o}},\ and\ \bibinfo {author} {\bibfnamefont {F.}~\bibnamefont {Winter}}
  (\bibinfo {collaboration} {NME}),\ }\bibfield  {title} {\bibinfo {title}
  {{Precision nucleon charges and form factors using (2+1)-flavor lattice
  QCD}},\ }\href {https://doi.org/10.1103/PhysRevD.105.054505} {\bibfield
  {journal} {\bibinfo  {journal} {Phys. Rev. D}\ }\textbf {\bibinfo {volume}
  {105}},\ \bibinfo {pages} {054505} (\bibinfo {year} {2022})},\ \Eprint
  {https://arxiv.org/abs/2103.05599} {arXiv:2103.05599 [hep-lat]} \BibitemShut
  {NoStop}%
\bibitem [{\citenamefont {Djukanovic}\ \emph {et~al.}(2022)\citenamefont
  {Djukanovic}, \citenamefont {von Hippel}, \citenamefont {Koponen},
  \citenamefont {Meyer}, \citenamefont {Ottnad}, \citenamefont {Schulz},\ and\
  \citenamefont {Wittig}}]{Djukanovic:2022wru}%
  \BibitemOpen
  \bibfield  {author} {\bibinfo {author} {\bibfnamefont {D.}~\bibnamefont
  {Djukanovic}}, \bibinfo {author} {\bibfnamefont {G.}~\bibnamefont {von
  Hippel}}, \bibinfo {author} {\bibfnamefont {J.}~\bibnamefont {Koponen}},
  \bibinfo {author} {\bibfnamefont {H.~B.}\ \bibnamefont {Meyer}}, \bibinfo
  {author} {\bibfnamefont {K.}~\bibnamefont {Ottnad}}, \bibinfo {author}
  {\bibfnamefont {T.}~\bibnamefont {Schulz}},\ and\ \bibinfo {author}
  {\bibfnamefont {H.}~\bibnamefont {Wittig}},\ }\bibfield  {title} {\bibinfo
  {title} {{Isovector axial form factor of the nucleon from lattice QCD}},\
  }\href {https://doi.org/10.1103/PhysRevD.106.074503} {\bibfield  {journal}
  {\bibinfo  {journal} {Phys. Rev. D}\ }\textbf {\bibinfo {volume} {106}},\
  \bibinfo {pages} {074503} (\bibinfo {year} {2022})},\ \Eprint
  {https://arxiv.org/abs/2207.03440} {arXiv:2207.03440 [hep-lat]} \BibitemShut
  {NoStop}%
\bibitem [{\citenamefont {Meyer}\ \emph {et~al.}(2016)\citenamefont {Meyer},
  \citenamefont {Betancourt}, \citenamefont {Gran},\ and\ \citenamefont
  {Hill}}]{Meyer:2016oeg}%
  \BibitemOpen
  \bibfield  {author} {\bibinfo {author} {\bibfnamefont {A.~S.}\ \bibnamefont
  {Meyer}}, \bibinfo {author} {\bibfnamefont {M.}~\bibnamefont {Betancourt}},
  \bibinfo {author} {\bibfnamefont {R.}~\bibnamefont {Gran}},\ and\ \bibinfo
  {author} {\bibfnamefont {R.~J.}\ \bibnamefont {Hill}},\ }\bibfield  {title}
  {\bibinfo {title} {{Deuterium target data for precision neutrino-nucleus
  cross sections}},\ }\href {https://doi.org/10.1103/PhysRevD.93.113015}
  {\bibfield  {journal} {\bibinfo  {journal} {Phys. Rev. D}\ }\textbf {\bibinfo
  {volume} {93}},\ \bibinfo {pages} {113015} (\bibinfo {year} {2016})},\
  \Eprint {https://arxiv.org/abs/1603.03048} {arXiv:1603.03048 [hep-ph]}
  \BibitemShut {NoStop}%
\bibitem [{\citenamefont {Bali}\ \emph {et~al.}(2015)\citenamefont {Bali},
  \citenamefont {Collins}, \citenamefont {Gl\"assle}, \citenamefont
  {G\"ockeler}, \citenamefont {Najjar}, \citenamefont {R\"odl}, \citenamefont
  {Sch\"afer}, \citenamefont {Schiel}, \citenamefont {S\"oldner},\ and\
  \citenamefont {Sternbeck}}]{Bali:2014nma}%
  \BibitemOpen
  \bibfield  {author} {\bibinfo {author} {\bibfnamefont {G.~S.}\ \bibnamefont
  {Bali}}, \bibinfo {author} {\bibfnamefont {S.}~\bibnamefont {Collins}},
  \bibinfo {author} {\bibfnamefont {B.}~\bibnamefont {Gl\"assle}}, \bibinfo
  {author} {\bibfnamefont {M.}~\bibnamefont {G\"ockeler}}, \bibinfo {author}
  {\bibfnamefont {J.}~\bibnamefont {Najjar}}, \bibinfo {author} {\bibfnamefont
  {R.~H.}\ \bibnamefont {R\"odl}}, \bibinfo {author} {\bibfnamefont
  {A.}~\bibnamefont {Sch\"afer}}, \bibinfo {author} {\bibfnamefont {R.~W.}\
  \bibnamefont {Schiel}}, \bibinfo {author} {\bibfnamefont {W.}~\bibnamefont
  {S\"oldner}},\ and\ \bibinfo {author} {\bibfnamefont {A.}~\bibnamefont
  {Sternbeck}} (\bibinfo {collaboration} {RQCD}),\ }\bibfield  {title}
  {\bibinfo {title} {{Nucleon isovector couplings from $N_f=2$ lattice QCD}},\
  }\href {https://doi.org/10.1103/PhysRevD.91.054501} {\bibfield  {journal}
  {\bibinfo  {journal} {Phys. Rev. D}\ }\textbf {\bibinfo {volume} {91}},\
  \bibinfo {pages} {054501} (\bibinfo {year} {2015})},\ \Eprint
  {https://arxiv.org/abs/1412.7336} {arXiv:1412.7336 [hep-lat]} \BibitemShut
  {NoStop}%
\bibitem [{\citenamefont {Capitani}\ \emph {et~al.}(2019)\citenamefont
  {Capitani}, \citenamefont {Della~Morte}, \citenamefont {Djukanovic},
  \citenamefont {von Hippel}, \citenamefont {Hua}, \citenamefont {J\"ager},
  \citenamefont {Junnarkar}, \citenamefont {Meyer}, \citenamefont {Rae},\ and\
  \citenamefont {Wittig}}]{Capitani:2017qpc}%
  \BibitemOpen
  \bibfield  {author} {\bibinfo {author} {\bibfnamefont {S.}~\bibnamefont
  {Capitani}}, \bibinfo {author} {\bibfnamefont {M.}~\bibnamefont
  {Della~Morte}}, \bibinfo {author} {\bibfnamefont {D.}~\bibnamefont
  {Djukanovic}}, \bibinfo {author} {\bibfnamefont {G.~M.}\ \bibnamefont {von
  Hippel}}, \bibinfo {author} {\bibfnamefont {J.}~\bibnamefont {Hua}}, \bibinfo
  {author} {\bibfnamefont {B.}~\bibnamefont {J\"ager}}, \bibinfo {author}
  {\bibfnamefont {P.~M.}\ \bibnamefont {Junnarkar}}, \bibinfo {author}
  {\bibfnamefont {H.~B.}\ \bibnamefont {Meyer}}, \bibinfo {author}
  {\bibfnamefont {T.~D.}\ \bibnamefont {Rae}},\ and\ \bibinfo {author}
  {\bibfnamefont {H.}~\bibnamefont {Wittig}},\ }\bibfield  {title} {\bibinfo
  {title} {{Isovector axial form factors of the nucleon in two-flavor lattice
  QCD}},\ }\href {https://doi.org/10.1142/S0217751X1950009X} {\bibfield
  {journal} {\bibinfo  {journal} {Int. J. Mod. Phys. A}\ }\textbf {\bibinfo
  {volume} {34}},\ \bibinfo {pages} {1950009} (\bibinfo {year} {2019})},\
  \Eprint {https://arxiv.org/abs/1705.06186} {arXiv:1705.06186 [hep-lat]}
  \BibitemShut {NoStop}%
\bibitem [{\citenamefont {Gupta}\ \emph {et~al.}(2017)\citenamefont {Gupta},
  \citenamefont {Jang}, \citenamefont {Lin}, \citenamefont {Yoon},\ and\
  \citenamefont {Bhattacharya}}]{Gupta:2017dwj}%
  \BibitemOpen
  \bibfield  {author} {\bibinfo {author} {\bibfnamefont {R.}~\bibnamefont
  {Gupta}}, \bibinfo {author} {\bibfnamefont {Y.-C.}\ \bibnamefont {Jang}},
  \bibinfo {author} {\bibfnamefont {H.-W.}\ \bibnamefont {Lin}}, \bibinfo
  {author} {\bibfnamefont {B.}~\bibnamefont {Yoon}},\ and\ \bibinfo {author}
  {\bibfnamefont {T.}~\bibnamefont {Bhattacharya}} (\bibinfo {collaboration}
  {PNDME}),\ }\bibfield  {title} {\bibinfo {title} {{Axial vector form factors
  of the nucleon from lattice QCD}},\ }\href
  {https://doi.org/10.1103/PhysRevD.96.114503} {\bibfield  {journal} {\bibinfo
  {journal} {Phys. Rev. D}\ }\textbf {\bibinfo {volume} {96}},\ \bibinfo
  {pages} {114503} (\bibinfo {year} {2017})},\ \Eprint
  {https://arxiv.org/abs/1705.06834} {arXiv:1705.06834 [hep-lat]} \BibitemShut
  {NoStop}%
\bibitem [{\citenamefont {Tsukamoto}\ \emph {et~al.}(2018)\citenamefont
  {Tsukamoto}, \citenamefont {Ishikawa}, \citenamefont {Kuramashi},
  \citenamefont {Sasaki},\ and\ \citenamefont {Yamazaki}}]{Tsukamoto:2017fnm}%
  \BibitemOpen
  \bibfield  {author} {\bibinfo {author} {\bibfnamefont {N.}~\bibnamefont
  {Tsukamoto}}, \bibinfo {author} {\bibfnamefont {K.-I.}\ \bibnamefont
  {Ishikawa}}, \bibinfo {author} {\bibfnamefont {Y.}~\bibnamefont {Kuramashi}},
  \bibinfo {author} {\bibfnamefont {S.}~\bibnamefont {Sasaki}},\ and\ \bibinfo
  {author} {\bibfnamefont {T.}~\bibnamefont {Yamazaki}} (\bibinfo
  {collaboration} {PACS}),\ }\bibfield  {title} {\bibinfo {title} {{Nucleon
  structure from $2+1$ flavor lattice QCD near the physical point}},\ }\href
  {https://doi.org/10.1051/epjconf/201817506007} {\bibfield  {journal}
  {\bibinfo  {journal} {EPJ Web Conf.}\ }\textbf {\bibinfo {volume} {175}},\
  \bibinfo {pages} {06007} (\bibinfo {year} {2018})},\ \Eprint
  {https://arxiv.org/abs/1710.10782} {arXiv:1710.10782 [hep-lat]} \BibitemShut
  {NoStop}%
\bibitem [{\citenamefont {Bali}\ \emph {et~al.}(2019)\citenamefont {Bali},
  \citenamefont {Collins}, \citenamefont {Gruber}, \citenamefont {Sch\"afer},
  \citenamefont {Wein},\ and\ \citenamefont {Wurm}}]{Bali:2018qus}%
  \BibitemOpen
  \bibfield  {author} {\bibinfo {author} {\bibfnamefont {G.~S.}\ \bibnamefont
  {Bali}}, \bibinfo {author} {\bibfnamefont {S.}~\bibnamefont {Collins}},
  \bibinfo {author} {\bibfnamefont {M.}~\bibnamefont {Gruber}}, \bibinfo
  {author} {\bibfnamefont {A.}~\bibnamefont {Sch\"afer}}, \bibinfo {author}
  {\bibfnamefont {P.}~\bibnamefont {Wein}},\ and\ \bibinfo {author}
  {\bibfnamefont {T.}~\bibnamefont {Wurm}} (\bibinfo {collaboration} {RQCD}),\
  }\bibfield  {title} {\bibinfo {title} {{Solving the PCAC puzzle for nucleon
  axial and pseudoscalar form factors}},\ }\href
  {https://doi.org/10.1016/j.physletb.2018.12.053} {\bibfield  {journal}
  {\bibinfo  {journal} {Phys. Lett. B}\ }\textbf {\bibinfo {volume} {789}},\
  \bibinfo {pages} {666} (\bibinfo {year} {2019})},\ \Eprint
  {https://arxiv.org/abs/1810.05569} {arXiv:1810.05569 [hep-lat]} \BibitemShut
  {NoStop}%
\bibitem [{\citenamefont {Bär}(2019{\natexlab{a}})}]{Bar:2018xyi}%
  \BibitemOpen
  \bibfield  {author} {\bibinfo {author} {\bibfnamefont {O.}~\bibnamefont
  {Bär}},\ }\bibfield  {title} {\bibinfo {title} {{$N\pi$-state contamination
  in lattice calculations of the nucleon axial form factors}},\ }\href
  {https://doi.org/10.1103/PhysRevD.99.054506} {\bibfield  {journal} {\bibinfo
  {journal} {Phys. Rev. D}\ }\textbf {\bibinfo {volume} {99}},\ \bibinfo
  {pages} {054506} (\bibinfo {year} {2019}{\natexlab{a}})},\ \Eprint
  {https://arxiv.org/abs/1812.09191} {arXiv:1812.09191 [hep-lat]} \BibitemShut
  {NoStop}%
\bibitem [{\citenamefont {Bär}(2019{\natexlab{b}})}]{Bar:2019gfx}%
  \BibitemOpen
  \bibfield  {author} {\bibinfo {author} {\bibfnamefont {O.}~\bibnamefont
  {Bär}},\ }\bibfield  {title} {\bibinfo {title} {{$N\pi$-state contamination
  in lattice calculations of the nucleon pseudoscalar form factor}},\ }\href
  {https://doi.org/10.1103/PhysRevD.100.054507} {\bibfield  {journal} {\bibinfo
   {journal} {Phys. Rev. D}\ }\textbf {\bibinfo {volume} {100}},\ \bibinfo
  {pages} {054507} (\bibinfo {year} {2019}{\natexlab{b}})},\ \Eprint
  {https://arxiv.org/abs/1906.03652} {arXiv:1906.03652 [hep-lat]} \BibitemShut
  {NoStop}%
\bibitem [{\citenamefont {Bär}(2020)}]{Bar:2019igf}%
  \BibitemOpen
  \bibfield  {author} {\bibinfo {author} {\bibfnamefont {O.}~\bibnamefont
  {Bär}},\ }\bibfield  {title} {\bibinfo {title} {{$N\pi$ states and the
  projection method for the nucleon axial and pseudoscalar form factors}},\
  }\href {https://doi.org/10.1103/PhysRevD.101.034515} {\bibfield  {journal}
  {\bibinfo  {journal} {Phys. Rev. D}\ }\textbf {\bibinfo {volume} {101}},\
  \bibinfo {pages} {034515} (\bibinfo {year} {2020})},\ \Eprint
  {https://arxiv.org/abs/1912.05873} {arXiv:1912.05873 [hep-lat]} \BibitemShut
  {NoStop}%
\bibitem [{\citenamefont {Tiburzi}(2015)}]{Tiburzi:2015tta}%
  \BibitemOpen
  \bibfield  {author} {\bibinfo {author} {\bibfnamefont {B.~C.}\ \bibnamefont
  {Tiburzi}},\ }\bibfield  {title} {\bibinfo {title} {{Chiral corrections to
  nucleon two- and three-point correlation functions}},\ }\href
  {https://doi.org/10.1103/PhysRevD.91.094510} {\bibfield  {journal} {\bibinfo
  {journal} {Phys. Rev. D}\ }\textbf {\bibinfo {volume} {91}},\ \bibinfo
  {pages} {094510} (\bibinfo {year} {2015})},\ \Eprint
  {https://arxiv.org/abs/1503.06329} {arXiv:1503.06329 [hep-lat]} \BibitemShut
  {NoStop}%
\bibitem [{\citenamefont {Hansen}\ and\ \citenamefont
  {Meyer}(2017)}]{Hansen:2016qoz}%
  \BibitemOpen
  \bibfield  {author} {\bibinfo {author} {\bibfnamefont {M.~T.}\ \bibnamefont
  {Hansen}}\ and\ \bibinfo {author} {\bibfnamefont {H.~B.}\ \bibnamefont
  {Meyer}},\ }\bibfield  {title} {\bibinfo {title} {{On the effect of excited
  states in lattice calculations of the nucleon axial charge}},\ }\href
  {https://doi.org/10.1016/j.nuclphysb.2017.08.017} {\bibfield  {journal}
  {\bibinfo  {journal} {Nucl. Phys. B}\ }\textbf {\bibinfo {volume} {923}},\
  \bibinfo {pages} {558} (\bibinfo {year} {2017})},\ \Eprint
  {https://arxiv.org/abs/1610.03843} {arXiv:1610.03843 [hep-lat]} \BibitemShut
  {NoStop}%
\bibitem [{\citenamefont {Ruso}\ \emph {et~al.}(2022)\citenamefont {Ruso} \emph
  {et~al.}}]{Ruso:2022qes}%
  \BibitemOpen
  \bibfield  {author} {\bibinfo {author} {\bibfnamefont {L.~A.}\ \bibnamefont
  {Ruso}} \emph {et~al.},\ }\bibfield  {title} {\bibinfo {title} {{Theoretical
  tools for neutrino scattering: interplay between lattice QCD, EFTs, nuclear
  physics, phenomenology, and neutrino event generators}},\ }\href@noop {} {\
  (\bibinfo {year} {2022})},\ \Eprint {https://arxiv.org/abs/2203.09030}
  {arXiv:2203.09030 [hep-ph]} \BibitemShut {NoStop}%
\bibitem [{\citenamefont {Göckeler}\ \emph {et~al.}(2012)\citenamefont
  {Göckeler}, \citenamefont {Horsley}, \citenamefont {Lage}, \citenamefont
  {Meißner}, \citenamefont {Rakow}, \citenamefont {Rusetsky}, \citenamefont
  {Schierholz},\ and\ \citenamefont {Zanotti}}]{Gockeler:2012yj}%
  \BibitemOpen
  \bibfield  {author} {\bibinfo {author} {\bibfnamefont {M.}~\bibnamefont
  {Göckeler}}, \bibinfo {author} {\bibfnamefont {R.}~\bibnamefont {Horsley}},
  \bibinfo {author} {\bibfnamefont {M.}~\bibnamefont {Lage}}, \bibinfo {author}
  {\bibfnamefont {U.}~\bibnamefont {Meißner}}, \bibinfo {author}
  {\bibfnamefont {P.~E.~L.}\ \bibnamefont {Rakow}}, \bibinfo {author}
  {\bibfnamefont {A.}~\bibnamefont {Rusetsky}}, \bibinfo {author}
  {\bibfnamefont {G.}~\bibnamefont {Schierholz}},\ and\ \bibinfo {author}
  {\bibfnamefont {J.~M.}\ \bibnamefont {Zanotti}},\ }\bibfield  {title}
  {\bibinfo {title} {{Scattering phases for meson and baryon resonances on
  general moving-frame lattices}},\ }\href
  {https://doi.org/10.1103/PhysRevD.86.094513} {\bibfield  {journal} {\bibinfo
  {journal} {Phys. Rev. D}\ }\textbf {\bibinfo {volume} {86}},\ \bibinfo
  {pages} {094513} (\bibinfo {year} {2012})},\ \Eprint
  {https://arxiv.org/abs/1206.4141} {arXiv:1206.4141 [hep-lat]} \BibitemShut
  {NoStop}%
\bibitem [{\citenamefont {Lang}\ \emph {et~al.}(2017)\citenamefont {Lang},
  \citenamefont {Leskovec}, \citenamefont {Padmanath},\ and\ \citenamefont
  {Prelovsek}}]{Lang:2016hnn}%
  \BibitemOpen
  \bibfield  {author} {\bibinfo {author} {\bibfnamefont {C.~B.}\ \bibnamefont
  {Lang}}, \bibinfo {author} {\bibfnamefont {L.}~\bibnamefont {Leskovec}},
  \bibinfo {author} {\bibfnamefont {M.}~\bibnamefont {Padmanath}},\ and\
  \bibinfo {author} {\bibfnamefont {S.}~\bibnamefont {Prelovsek}},\ }\bibfield
  {title} {\bibinfo {title} {{Pion-nucleon scattering in the {Roper} channel
  from lattice QCD}},\ }\href {https://doi.org/10.1103/PhysRevD.95.014510}
  {\bibfield  {journal} {\bibinfo  {journal} {Phys. Rev. D}\ }\textbf {\bibinfo
  {volume} {95}},\ \bibinfo {pages} {014510} (\bibinfo {year} {2017})},\
  \Eprint {https://arxiv.org/abs/1610.01422} {arXiv:1610.01422 [hep-lat]}
  \BibitemShut {NoStop}%
\bibitem [{\citenamefont {Prelovsek}\ \emph {et~al.}(2017)\citenamefont
  {Prelovsek}, \citenamefont {Skerbis},\ and\ \citenamefont
  {Lang}}]{Prelovsek:2016iyo}%
  \BibitemOpen
  \bibfield  {author} {\bibinfo {author} {\bibfnamefont {S.}~\bibnamefont
  {Prelovsek}}, \bibinfo {author} {\bibfnamefont {U.}~\bibnamefont {Skerbis}},\
  and\ \bibinfo {author} {\bibfnamefont {C.~B.}\ \bibnamefont {Lang}},\
  }\bibfield  {title} {\bibinfo {title} {{Lattice operators for scattering of
  particles with spin}},\ }\href {https://doi.org/10.1007/JHEP01(2017)129}
  {\bibfield  {journal} {\bibinfo  {journal} {J. High Energy Phys.}\ }\textbf
  {\bibinfo {volume} {01}},\ \bibinfo {pages} {129}},\ \Eprint
  {https://arxiv.org/abs/1607.06738} {arXiv:1607.06738 [hep-lat]} \BibitemShut
  {NoStop}%
\bibitem [{\citenamefont {Maiani}\ \emph {et~al.}(1987)\citenamefont {Maiani},
  \citenamefont {Martinelli}, \citenamefont {Paciello},\ and\ \citenamefont
  {Taglienti}}]{Maiani:1987by}%
  \BibitemOpen
  \bibfield  {author} {\bibinfo {author} {\bibfnamefont {L.}~\bibnamefont
  {Maiani}}, \bibinfo {author} {\bibfnamefont {G.}~\bibnamefont {Martinelli}},
  \bibinfo {author} {\bibfnamefont {M.~L.}\ \bibnamefont {Paciello}},\ and\
  \bibinfo {author} {\bibfnamefont {B.}~\bibnamefont {Taglienti}},\ }\bibfield
  {title} {\bibinfo {title} {{Scalar densities and baryon mass differences in
  lattice {QCD} with Wilson fermions}},\ }\href
  {https://doi.org/10.1016/0550-3213(87)90078-2} {\bibfield  {journal}
  {\bibinfo  {journal} {Nucl. Phys. B}\ }\textbf {\bibinfo {volume} {293}},\
  \bibinfo {pages} {420} (\bibinfo {year} {1987})}\BibitemShut {NoStop}%
\bibitem [{\citenamefont {Sommer}(1995)}]{Sommer:1994gg}%
  \BibitemOpen
  \bibfield  {author} {\bibinfo {author} {\bibfnamefont {R.}~\bibnamefont
  {Sommer}},\ }\bibfield  {title} {\bibinfo {title} {{Leptonic decays of $B$
  and $D$ mesons}},\ }\bibfield  {booktitle} {\emph {\bibinfo {booktitle}
  {{Proceedings, 12th International Symposium on Lattice Field Theory (Lattice
  94), Bielefeld, Germany, September 27--October 1, 1994}}},\ }\href
  {https://doi.org/10.1016/0920-5632(95)00201-J} {\bibfield  {journal}
  {\bibinfo  {journal} {Nucl. Phys. Proc. Suppl.}\ }\textbf {\bibinfo {volume}
  {42}},\ \bibinfo {pages} {186} (\bibinfo {year} {1995})},\ \Eprint
  {https://arxiv.org/abs/hep-lat/9411024} {arXiv:hep-lat/9411024 [hep-lat]}
  \BibitemShut {NoStop}%
\bibitem [{\citenamefont {Foster}\ and\ \citenamefont
  {Michael}(1999{\natexlab{a}})}]{Foster:1998wu}%
  \BibitemOpen
  \bibfield  {author} {\bibinfo {author} {\bibfnamefont {M.}~\bibnamefont
  {Foster}}\ and\ \bibinfo {author} {\bibfnamefont {C.}~\bibnamefont {Michael}}
  (\bibinfo {collaboration} {UKQCD}),\ }\bibfield  {title} {\bibinfo {title}
  {{Hadrons with a heavy color adjoint particle}},\ }\href
  {https://doi.org/10.1103/PhysRevD.59.094509} {\bibfield  {journal} {\bibinfo
  {journal} {Phys. Rev. D}\ }\textbf {\bibinfo {volume} {59}},\ \bibinfo
  {pages} {094509} (\bibinfo {year} {1999}{\natexlab{a}})},\ \Eprint
  {https://arxiv.org/abs/hep-lat/9811010} {arXiv:hep-lat/9811010 [hep-lat]}
  \BibitemShut {NoStop}%
\bibitem [{\citenamefont {Foster}\ and\ \citenamefont
  {Michael}(1999{\natexlab{b}})}]{Foster:1999oet}%
  \BibitemOpen
  \bibfield  {author} {\bibinfo {author} {\bibfnamefont {M.}~\bibnamefont
  {Foster}}\ and\ \bibinfo {author} {\bibfnamefont {C.}~\bibnamefont {Michael}}
  (\bibinfo {collaboration} {UKQCD Collaboration}),\ }\bibfield  {title}
  {\bibinfo {title} {Quark mass dependence of hadron masses from lattice
  {QCD}},\ }\href {https://doi.org/10.1103/PhysRevD.59.074503} {\bibfield
  {journal} {\bibinfo  {journal} {Phys. Rev. D}\ }\textbf {\bibinfo {volume}
  {59}},\ \bibinfo {pages} {074503} (\bibinfo {year}
  {1999}{\natexlab{b}})}\BibitemShut {NoStop}%
\bibitem [{\citenamefont {Bruno}\ \emph {et~al.}(2015)\citenamefont {Bruno}
  \emph {et~al.}}]{Bruno:2014jqa}%
  \BibitemOpen
  \bibfield  {author} {\bibinfo {author} {\bibfnamefont {M.}~\bibnamefont
  {Bruno}} \emph {et~al.} (\bibinfo {collaboration} {CLS}),\ }\bibfield
  {title} {\bibinfo {title} {{Simulation of QCD with N$_{f} =$ 2 $+$ 1 flavors
  of non-perturbatively improved Wilson fermions}},\ }\href
  {https://doi.org/10.1007/JHEP02(2015)043} {\bibfield  {journal} {\bibinfo
  {journal} {J. High Energy Phys.}\ }\textbf {\bibinfo {volume} {02}},\
  \bibinfo {pages} {043}},\ \Eprint {https://arxiv.org/abs/1411.3982}
  {arXiv:1411.3982 [hep-lat]} \BibitemShut {NoStop}%
\bibitem [{\citenamefont {Bali}\ \emph {et~al.}(2022)\citenamefont {Bali},
  \citenamefont {Collins}, \citenamefont {Georg}, \citenamefont {Jenkins},
  \citenamefont {Korcyl}, \citenamefont {Sch\"afer}, \citenamefont {Scholz},
  \citenamefont {Simeth}, \citenamefont {S\"oldner},\ and\ \citenamefont
  {Weish\"aupl}}]{RQCD:2022xux}%
  \BibitemOpen
  \bibfield  {author} {\bibinfo {author} {\bibfnamefont {G.~S.}\ \bibnamefont
  {Bali}}, \bibinfo {author} {\bibfnamefont {S.}~\bibnamefont {Collins}},
  \bibinfo {author} {\bibfnamefont {P.}~\bibnamefont {Georg}}, \bibinfo
  {author} {\bibfnamefont {D.}~\bibnamefont {Jenkins}}, \bibinfo {author}
  {\bibfnamefont {P.}~\bibnamefont {Korcyl}}, \bibinfo {author} {\bibfnamefont
  {A.}~\bibnamefont {Sch\"afer}}, \bibinfo {author} {\bibfnamefont {E.~E.}\
  \bibnamefont {Scholz}}, \bibinfo {author} {\bibfnamefont {J.}~\bibnamefont
  {Simeth}}, \bibinfo {author} {\bibfnamefont {W.}~\bibnamefont {S\"oldner}},\
  and\ \bibinfo {author} {\bibfnamefont {S.}~\bibnamefont {Weish\"aupl}}
  (\bibinfo {collaboration} {RQCD}),\ }\bibfield  {title} {\bibinfo {title}
  {{Scale setting and the light baryon spectrum in $N_f=2+1$ QCD with Wilson
  fermions}},\ }\href@noop {} {\  (\bibinfo {year} {2022})},\ \Eprint
  {https://arxiv.org/abs/2211.03744} {arXiv:2211.03744 [hep-lat]} \BibitemShut
  {NoStop}%
\bibitem [{\citenamefont {Berg}(1982)}]{Berg:1982hf}%
  \BibitemOpen
  \bibfield  {author} {\bibinfo {author} {\bibfnamefont {B.}~\bibnamefont
  {Berg}},\ }\bibfield  {title} {\bibinfo {title} {{Glueball calculations in
  lattice gauge theories}},\ }\href {https://doi.org/10.1051/jphyscol:1982355}
  {\bibfield  {journal} {\bibinfo  {journal} {J. Phys. Colloq.}\ }\textbf
  {\bibinfo {volume} {43}},\ \bibinfo {pages} {272} (\bibinfo {year}
  {1982})}\BibitemShut {NoStop}%
\bibitem [{\citenamefont {Michael}(1985)}]{Michael:1985ne}%
  \BibitemOpen
  \bibfield  {author} {\bibinfo {author} {\bibfnamefont {C.}~\bibnamefont
  {Michael}},\ }\bibfield  {title} {\bibinfo {title} {{Adjoint sources in
  lattice gauge theory}},\ }\href
  {https://doi.org/10.1016/0550-3213(85)90297-4} {\bibfield  {journal}
  {\bibinfo  {journal} {Nucl. Phys. B}\ }\textbf {\bibinfo {volume} {259}},\
  \bibinfo {pages} {58} (\bibinfo {year} {1985})}\BibitemShut {NoStop}%
\bibitem [{\citenamefont {Lüscher}\ and\ \citenamefont
  {Wolff}(1990)}]{Luscher:1990ck}%
  \BibitemOpen
  \bibfield  {author} {\bibinfo {author} {\bibfnamefont {M.}~\bibnamefont
  {Lüscher}}\ and\ \bibinfo {author} {\bibfnamefont {U.}~\bibnamefont
  {Wolff}},\ }\bibfield  {title} {\bibinfo {title} {{How to calculate the
  elastic scattering matrix in two-dimensional quantum field theories by
  numerical simulation}},\ }\href
  {https://doi.org/10.1016/0550-3213(90)90540-T} {\bibfield  {journal}
  {\bibinfo  {journal} {Nucl. Phys. B}\ }\textbf {\bibinfo {volume} {339}},\
  \bibinfo {pages} {222} (\bibinfo {year} {1990})}\BibitemShut {NoStop}%
\bibitem [{\citenamefont {Blossier}\ \emph {et~al.}(2009)\citenamefont
  {Blossier}, \citenamefont {Della~Morte}, \citenamefont {von Hippel},
  \citenamefont {Mendes},\ and\ \citenamefont {Sommer}}]{Blossier:2009kd}%
  \BibitemOpen
  \bibfield  {author} {\bibinfo {author} {\bibfnamefont {B.}~\bibnamefont
  {Blossier}}, \bibinfo {author} {\bibfnamefont {M.}~\bibnamefont
  {Della~Morte}}, \bibinfo {author} {\bibfnamefont {G.}~\bibnamefont {von
  Hippel}}, \bibinfo {author} {\bibfnamefont {T.}~\bibnamefont {Mendes}},\ and\
  \bibinfo {author} {\bibfnamefont {R.}~\bibnamefont {Sommer}},\ }\bibfield
  {title} {\bibinfo {title} {{On the generalized eigenvalue method for energies
  and matrix elements in lattice field theory}},\ }\href
  {https://doi.org/10.1088/1126-6708/2009/04/094} {\bibfield  {journal}
  {\bibinfo  {journal} {J. High Energy Phys.}\ }\textbf {\bibinfo {volume}
  {04}},\ \bibinfo {pages} {094}},\ \Eprint {https://arxiv.org/abs/0902.1265}
  {arXiv:0902.1265 [hep-lat]} \BibitemShut {NoStop}%
\bibitem [{\citenamefont {Bulava}\ \emph {et~al.}(2012)\citenamefont {Bulava},
  \citenamefont {Donnellan},\ and\ \citenamefont {Sommer}}]{Bulava:2011yz}%
  \BibitemOpen
  \bibfield  {author} {\bibinfo {author} {\bibfnamefont {J.}~\bibnamefont
  {Bulava}}, \bibinfo {author} {\bibfnamefont {M.}~\bibnamefont {Donnellan}},\
  and\ \bibinfo {author} {\bibfnamefont {R.}~\bibnamefont {Sommer}},\
  }\bibfield  {title} {\bibinfo {title} {{On the computation of
  hadron-to-hadron transition matrix elements in lattice QCD}},\ }\href
  {https://doi.org/10.1007/JHEP01(2012)140} {\bibfield  {journal} {\bibinfo
  {journal} {J. High Energy Phys.}\ }\textbf {\bibinfo {volume} {01}},\
  \bibinfo {pages} {140}},\ \Eprint {https://arxiv.org/abs/1108.3774}
  {arXiv:1108.3774 [hep-lat]} \BibitemShut {NoStop}%
\bibitem [{\citenamefont {Owen}\ \emph {et~al.}(2013)\citenamefont {Owen},
  \citenamefont {Dragos}, \citenamefont {Kamleh}, \citenamefont {Leinweber},
  \citenamefont {Mahbub}, \citenamefont {Menadue},\ and\ \citenamefont
  {Zanotti}}]{Owen:2012ts}%
  \BibitemOpen
  \bibfield  {author} {\bibinfo {author} {\bibfnamefont {B.~J.}\ \bibnamefont
  {Owen}}, \bibinfo {author} {\bibfnamefont {J.}~\bibnamefont {Dragos}},
  \bibinfo {author} {\bibfnamefont {W.}~\bibnamefont {Kamleh}}, \bibinfo
  {author} {\bibfnamefont {D.~B.}\ \bibnamefont {Leinweber}}, \bibinfo {author}
  {\bibfnamefont {M.~S.}\ \bibnamefont {Mahbub}}, \bibinfo {author}
  {\bibfnamefont {B.~J.}\ \bibnamefont {Menadue}},\ and\ \bibinfo {author}
  {\bibfnamefont {J.~M.}\ \bibnamefont {Zanotti}},\ }\bibfield  {title}
  {\bibinfo {title} {{Variational approach to the calculation of $g_A$}},\
  }\href {https://doi.org/10.1016/j.physletb.2013.04.063} {\bibfield  {journal}
  {\bibinfo  {journal} {Phys. Lett. B}\ }\textbf {\bibinfo {volume} {723}},\
  \bibinfo {pages} {217} (\bibinfo {year} {2013})},\ \Eprint
  {https://arxiv.org/abs/1212.4668} {arXiv:1212.4668 [hep-lat]} \BibitemShut
  {NoStop}%
\bibitem [{\citenamefont {Dragos}\ \emph {et~al.}(2016)\citenamefont {Dragos},
  \citenamefont {Horsley}, \citenamefont {Kamleh}, \citenamefont {Leinweber},
  \citenamefont {Nakamura}, \citenamefont {Rakow}, \citenamefont {Schierholz},
  \citenamefont {Young},\ and\ \citenamefont {Zanotti}}]{Dragos:2016rtx}%
  \BibitemOpen
  \bibfield  {author} {\bibinfo {author} {\bibfnamefont {J.}~\bibnamefont
  {Dragos}}, \bibinfo {author} {\bibfnamefont {R.}~\bibnamefont {Horsley}},
  \bibinfo {author} {\bibfnamefont {W.}~\bibnamefont {Kamleh}}, \bibinfo
  {author} {\bibfnamefont {D.~B.}\ \bibnamefont {Leinweber}}, \bibinfo {author}
  {\bibfnamefont {Y.}~\bibnamefont {Nakamura}}, \bibinfo {author}
  {\bibfnamefont {P.~E.~L.}\ \bibnamefont {Rakow}}, \bibinfo {author}
  {\bibfnamefont {G.}~\bibnamefont {Schierholz}}, \bibinfo {author}
  {\bibfnamefont {R.~D.}\ \bibnamefont {Young}},\ and\ \bibinfo {author}
  {\bibfnamefont {J.~M.}\ \bibnamefont {Zanotti}},\ }\bibfield  {title}
  {\bibinfo {title} {{Nucleon matrix elements using the variational method in
  lattice QCD}},\ }\href {https://doi.org/10.1103/PhysRevD.94.074505}
  {\bibfield  {journal} {\bibinfo  {journal} {Phys. Rev. D}\ }\textbf {\bibinfo
  {volume} {94}},\ \bibinfo {pages} {074505} (\bibinfo {year} {2016})},\
  \Eprint {https://arxiv.org/abs/1606.03195} {arXiv:1606.03195 [hep-lat]}
  \BibitemShut {NoStop}%
\bibitem [{\citenamefont {Liang}\ \emph {et~al.}(2017)\citenamefont {Liang},
  \citenamefont {Yang}, \citenamefont {Liu}, \citenamefont {Alexandru},
  \citenamefont {Draper},\ and\ \citenamefont {Sufian}}]{Liang:2016fgy}%
  \BibitemOpen
  \bibfield  {author} {\bibinfo {author} {\bibfnamefont {J.}~\bibnamefont
  {Liang}}, \bibinfo {author} {\bibfnamefont {Y.-B.}\ \bibnamefont {Yang}},
  \bibinfo {author} {\bibfnamefont {K.-F.}\ \bibnamefont {Liu}}, \bibinfo
  {author} {\bibfnamefont {A.}~\bibnamefont {Alexandru}}, \bibinfo {author}
  {\bibfnamefont {T.}~\bibnamefont {Draper}},\ and\ \bibinfo {author}
  {\bibfnamefont {R.~S.}\ \bibnamefont {Sufian}} (\bibinfo {collaboration}
  {$\chi$QCD}),\ }\bibfield  {title} {\bibinfo {title} {{Lattice calculation of
  nucleon isovector axial charge with improved currents}},\ }\href
  {https://doi.org/10.1103/PhysRevD.96.034519} {\bibfield  {journal} {\bibinfo
  {journal} {Phys. Rev. D}\ }\textbf {\bibinfo {volume} {96}},\ \bibinfo
  {pages} {034519} (\bibinfo {year} {2017})},\ \Eprint
  {https://arxiv.org/abs/1612.04388} {arXiv:1612.04388 [hep-lat]} \BibitemShut
  {NoStop}%
\bibitem [{\citenamefont {{J\"{u}lich Supercomputing Centre}}(2018)}]{jureca}%
  \BibitemOpen
  \bibfield  {author} {\bibinfo {author} {\bibnamefont {{J\"{u}lich
  Supercomputing Centre}}},\ }\bibfield  {title} {\bibinfo {title} {{JURECA:
  modular supercomputer at J\"{u}lich Supercomputing Centre}},\ }\href
  {https://doi.org/10.17815/jlsrf-4-121-1} {\bibfield  {journal} {\bibinfo
  {journal} {J. of large-scale research facilities}\ }\textbf {\bibinfo
  {volume} {4}},\ \bibinfo {pages} {A132} (\bibinfo {year} {2018})}\BibitemShut
  {NoStop}%
\bibitem [{\citenamefont {Edwards}\ and\ \citenamefont
  {Jo\'o}(2005)}]{Edwards:2004sx}%
  \BibitemOpen
  \bibfield  {author} {\bibinfo {author} {\bibfnamefont {R.~G.}\ \bibnamefont
  {Edwards}}\ and\ \bibinfo {author} {\bibfnamefont {B.}~\bibnamefont {Jo\'o}}
  (\bibinfo {collaboration} {SciDAC, LHP and UKQCD}),\ }\bibfield  {title}
  {\bibinfo {title} {{The Chroma software system for Lattice QCD}},\ }\href
  {https://doi.org/10.1016/j.nuclphysbps.2004.11.254} {\bibfield  {journal}
  {\bibinfo  {journal} {Nucl. Phys. B Proc. Suppl.}\ }\textbf {\bibinfo
  {volume} {140}},\ \bibinfo {pages} {832} (\bibinfo {year} {2005})},\ \Eprint
  {https://arxiv.org/abs/hep-lat/0409003} {arXiv:hep-lat/0409003 [hep-lat]}
  \BibitemShut {NoStop}%
\end{thebibliography}%

\end{document}